\documentclass[aps,prx,floatfix,twocolumn,showpacs,10pt,longbibliography]{revtex4-1}
\usepackage{amsmath}
\usepackage{url}
\usepackage{graphicx}
\usepackage{dcolumn}
\usepackage{scrextend}
\usepackage{hyperref}
\usepackage{bm}
\usepackage{amssymb}
\usepackage{rotating}
\usepackage[abs]{overpic}
\usepackage{xcolor}
\usepackage{tabularx}
\usepackage{floatrow}
\usepackage{subfig} 
\usepackage[justification=RaggedRight]{caption}
\usepackage{hyperref}
\hypersetup{colorlinks = true, citebordercolor={blue}, linkcolor={blue}, citecolor={blue}, urlcolor={blue}}
\begin{document}




\title{Preparation of an Exciton Condensate of Photons on a 53-Qubit Quantum Computer}

\author{LeeAnn M. Sager, Scott E. Smart, and David A. Mazziotti}

\email{damazz@uchicago.edu}

\affiliation{Department of Chemistry and The James Franck Institute, The University of Chicago, Chicago, IL 60637}%

\date{Submitted April 28, 2020}






\begin{abstract}

Quantum computation promises an exponential speedup of certain classes of classical calculations through the preparation and manipulation of entangled quantum states.  So far most molecular simulations on quantum computers, however, have been limited to small numbers of particles.  Here we prepare a highly entangled state on a 53-qubit IBM quantum computer, representing 53 particles, which reveals the formation of an exciton condensate of photon particles and holes.  While elusive for more than 50 years, such condensates were recently achieved for electron-hole pairs in graphene bilayers and metal chalcogenides.  Our result with a photon condensate has the potential to further the exploration of this new form of condensate that may play a significant role in realizing efficient room-temperature energy transport.

\end{abstract}

\maketitle


\noindent  Exciton condensation is defined by the condensation of particle-hole pairs (excitons) into a single quantum state to create a superfluid.  The superfluidity of electron-hole pairs---while, by definition, incapable of involving either the frictionless flow of matter or charge---does involve the non-dissipative transfer of energy \cite{Fil_Shevchenko_Rev,keldysh_2017}. As such, understanding and exploiting  the superfluid properties of exciton condensates may be instrumental in the effort to design wires and electronic devices with minimal loss of energy.  Consequently, considerable theoretical and experimental investigation has centered on exciton condensation in recent years \cite{KSE2004,TSH2004,Fil_Shevchenko_Rev,Shiva,Kogar2017,LWT2017,varsano_2017}.

While excitons form spontaneously in semiconductors and insulators and while the binding energy of the excitons can greatly exceed their thermal energy at room temperature, they recombine too quickly to allow for formation of a condensate in a simple manner.  To combat recombination, the coupling of excitons to polaritons---which requires the continuous input of light \cite{KRK2006,fuhrer_hamilton_2016}---and the physical separation of electrons and holes into bilayers---which involves impractically high magnetic fields and/or low temperatures \cite{KSE2004,TSH2004,SEP2000,NFE2012,fuhrer_hamilton_2016}---are employed.  Thus, a new, more-practical avenue for the creation of exciton condensates and the study of their properties is desired.

Computation has recently been employed to explore strongly-correlated quantum matter \cite{schuster_2019}, as superconducting circuits allow for precise manipulation of the strongly-interacting qubits to create specified quantum states populated by microwave photons.  Here we prepare and measure an exciton condensate of photons on a quantum computer.    Quantum computation should be particularly adapted to the exploration of exciton condensation as the binary nature of an individual qubit can be interpreted as a site consisting of one fermion and two orbitals; extrapolating, a system of $N$ qubits can be viewed as $N$ degenerate sites each consisting of one fermion and two orbitals.  We use such a Hamiltonian of $N$-fermions in two $N$-degenerate levels, known as the Lipkin Hamiltonian \cite{Lipkin_model,texpansion,David1998,Stein_2000,David2004}, to prepare \color{black} a molecular-scale \color{black} exciton condensate by tuning its interaction parameter to the large-coupling limit.
Because each transmon qubit on the quantum computer \color{black} utilized for this study \color{black} employs a microwave photon in an anharmonic quantum well potential, the exciton condensates \color{black} we  construct  are \color{black} comprised of photon-hole pairs condensing into single quantum states---\color{black}i.e., \color{black} exciton condensates of photons on a quantum computer.

We use the theoretical signature of exciton condensation---derived by Rosina and Garrod \cite{GR1969,Shiva}---to probe the extent of exciton condensation for a wide range of preparations through simulation and physical quantum computation experiments.  From analysis of the natural occupation numbers of these preparations, we establish that \color{black} prepared states with orbital occupation numbers consistent with the Greenberger-Horne-Zeilinger (GHZ) state\color{black}---including but not limited to the GHZ state---demonstrate maximal exciton condensation for three qubits.  Further, we establish \color{black} through simulation \color{black} that for any number of qubits, the GHZ state exhibits maximal character of exciton condensation, demonstrating that the ``maximal entanglement'' of the GHZ state---for all $N$---corresponds to the entanglement of particle-hole pairs.
\color{black} Through preparing and probing the GHZ states on quantum devices, character of exciton condensation is experimentally observed in systems composed of up to fifty-three qubits, although decoherence in the higher-qubit systems leads to multiple eigenstates demonstrating excitonic character instead of a single, maximally-entangled eigenstate.
\color{black}
Specifically as the GHZ state is prepared \color{black} here \color{black} on transmon qubits and is hence occupied by microwave photons, \color{black} exciton condensation of photons---\color{black}i.e., the entanglement of photon-hole pairs---is experimentally observed for systems of three to fifty-three qubits.
\color{black}

\vspace{0.1cm}
\noindent\textbf{\large{Results:}}

\noindent\textbf{Establishing signature of exciton condensation.}  Condensation phenomena has been an active area of research since 1924 when Einstein and Bose first introduced their ideal ``Bose-Einstein'' gas \cite{bose_einstein_1924,einstein_1924}. The identical particles comprising this gas---bosons---are able to aggregate into a single quantum ground state when sufficiently cooled \cite{einstein_1924}, which leads to superfluidity of the constituent bosons \cite{london_1938,tisza_1947}.  In 1940, Pauli established the relationship between spin and statistics, demonstrating that particles with integral spin values (bosons) obey Bose-Einstein statistics and hence may form a condensate \cite{pauli_1940}. Particles with half-integer spins (fermions), in contrast, must obey the Pauli exclusion principle and are therefore unable to condense into a single quantum state to form a condensate.  However, pairs of fermions---forming an overall bosonic state---can condense.  In a system of fermionic particles, this pairing can be accomplished through either particle-particle or particle-hole pairing.  The condensation of particle-particle pairs into a single quantum state is termed fermion-pair condensation with the resultant superfluidity of fermion pairs causing superconductivity \cite{Anderson_2013}; likewise, the condensation of particle-hole pairs (excitons) into a single quantum state is termed exciton condensation with the resultant superfluidity of exciton pairs causing the nondissipative transfer of energy \cite{Fil_Shevchenko_Rev}.

In order to \color{black} computationally probe \color{black} the presence and extent of condensation behavior, it is useful to establish a calculable, characteristic property.  As proven independently by Yang and Sasaki \cite{Y1962,S1965}, the quantum signature of fermion condensation is associated with a large eigenvalue of the particle-particle RDM with elements given by
\begin{equation}
^{2} D_{k,l}^{i,j} = \langle \Psi | {\hat a}^{\dagger}_i {\hat a}^{\dagger}_j {\hat a}_l {\hat a}_k  | \Psi \rangle
\label{eq:D2}
\end{equation}
where $|\Psi\rangle$ is an $N$-fermion wavefunction, the roman indices correspond to one-fermion orbitals in a finite basis set with rank $r$, and $\hat{a}^\dagger$ and $\hat{a}$ are fermionic creation and annihilation operators respectively.
We denote the largest eigenvalue of the particle-particle RDM as $\lambda_D$ and use this value as a signature of the extent of fermion pair condensation, with values above one demonstrating condensation.

In analogy to the signature of fermion pair (particle-particle) condensation being a large eigenvalue of the particle-particle RDM,  one may assume the quantum signature of exciton (particle-hole) condensation to be a large eigenvalue in the particle-hole RDM \cite{Shiva,GR1969,Kohn1970} with elements given by
\begin{equation}
^{2} G_{k,l}^{i,j} =  \langle \Psi | {\hat a}^{\dagger}_i {\hat a}_j {\hat a}^{\dagger}_l{\hat a}_k  | \Psi \rangle.
\label{G2}
\end{equation}
However, there exist two large eigenvalues for the particle-hole RDM, one of which corresponds to a ground-state-to-ground-state transition (not exciton condensation).  In order to eliminate this extraneous large eigenvalue, the modified particle-hole matrix with the ground-state resolution removed
\begin{equation}
^{2} {\tilde G}_{k,l}^{i,j} =^2G_{k,l}^{i,j} -{^{1}D_{k}^{i}} {^{1} D_{l}^{j}}
\label{G2tilde}
\end{equation}
is constructed.  Garrod and Rosina \cite{GR1969} have shown that---for an $N$-fermion system---the eigenvalues of the $^{2}\Tilde{G}$ matrix are zero or one in the non-interacting limit and bounded above by $\frac{N}{2}$ in the limit of strong correlation.  We denote the largest eigenvalue of the modified particle-hole RDM as $\lambda_G$ and use this value as a signature of the presence and extent of exciton condensation.

\noindent\textbf{Exploration of $\mathbf{\lambda_G}$ \color{black} for Three-Qubit Preparations.}
\noindent \color{black} Three-qubit systems---which correspond to three fermions in six orbitals---are the smallest systems to possess nontrivial classes of entanglement.  Hence, \color{black} in this study, \color{black} these minimally-small\color{black}, three-qubit \color{black} systems are first thoroughly explored in order to obtain insights on the preparation and characteristics of exciton condensates that are later employed to guide the investigation of larger-qubit systems.

To this end, the three-qubit preparation \color{black}
\begin{equation}
	|\Psi\rangle=C_3^2R_{y,3}(\theta_3)C_1^2R_{y,1}(\theta_2)C_1^3R_{y,1}(\theta_1)|000\rangle,
\label{eq:prep}
\end{equation}
\color{black}
which---as shown in Ref. \onlinecite{smart_2019}---is a minimalistic three-qubit preparation known to effectively span all real, 1-qubit occupations, is utilized to systematically prepare all real, three-qubit quantum states up to local unitaries.  Note that in this preparation, \color{black} $|000\rangle$ represents the initial all-zero qubit state, $C_i^j$ is a controlled-NOT (CNOT) gate with $i$ control and $j$ target, and $R_{y,i}(\theta)$ is a $\frac{\theta}{2}$ angle rotation about the y-axis of the Bloch sphere on the $i^{th}$ qubit.
\color{black}
By scanning over the angles of rotation ($\theta_1$, $\theta_2$, $\theta_3$), we prepare states with all possible, real qubit occupation numbers and hence sweep through all possible correlation phenomena.  By probing $\lambda_G$ for each of the prepared states, we then determine the extent of exciton condensation for all three-qubit correlation.
\color{black}

\begin{figure}[tbh!]
  \centering
  \sidesubfloat[Simulation]{\label{fig:sim_0}\includegraphics[scale=.40]{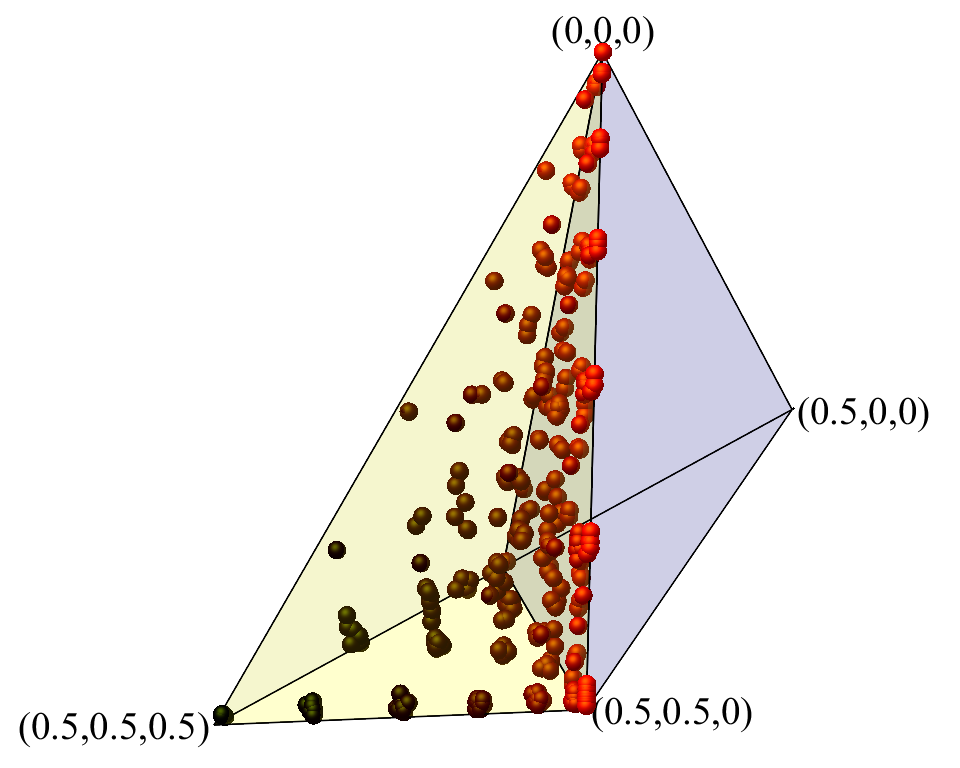}}\\
  \sidesubfloat[Mitigated]{\label{fig:exp_0}\includegraphics[scale=.40]{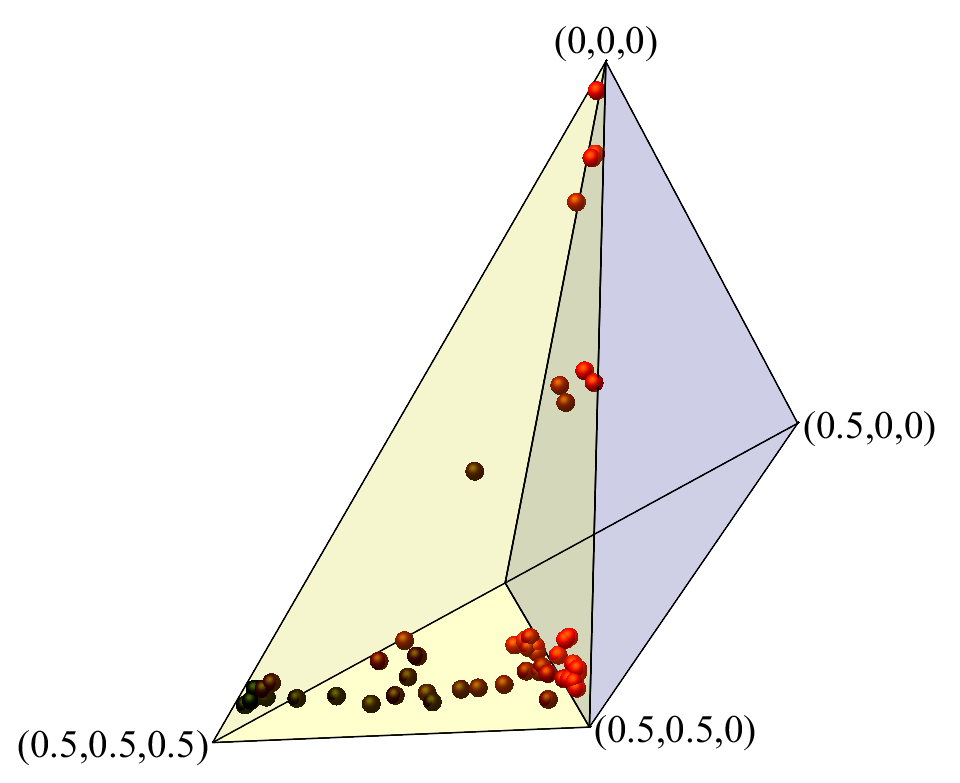}}\\
  \caption{\textbf{ Exciton Condensation in the Generalized Pauli Polytope.}  (a) Simulated and (b) experimental data shows that the occupation numbers ($n_{4}$, $n_{5}$, $n_{6}$) of the 1-RDM lie in the generalized Pauli polytope (yellow region) with exciton condensate character (darkening red indicates an increasing $\lambda_{G}$) emerging as the occupations saturate the vertex $(0.5, 0.5, 0.5)$.}
\label{fig:all_scan_poly}
\end{figure}

Orbital occupation numbers---obtained from the eigenvalues of the one-fermion RDM---are used as a practical coordinate representation in which to visualize  $\lambda_G$ for all electron correlations (all possible occupations).  For a three-qubit system, a pure quantum system of three electrons in six orbitals, these occupations are constrained beyond the traditional Pauli constraints ($0\le n_i \le 1$) \cite{borland_dennis_1972}.   For a three-qubit quantum system, \color{black} these \color{black} relevant so-called generalized Pauli constraints are \color{black}
\begin{equation}
	n_5+n_6-n_4 \ge 0
\label{eq:paulicons}
\end{equation}
where
\begin{gather}
	n_1+n_6=1 \\
	n_2+n_5=1 \\
	n_3+n_4=1
\label{eq:paulicons_more}
\end{gather}
in which each $n_i$ corresponds the natural-orbital occupations ordered from largest to smallest  \cite{schilling_2013,benavides_2013,chakraborty_2014,mazziotti_2016} .
\color{black}
The three, independent eigenvalues, $n_4$, $n_5$, and $n_6$, can be used as a three-coordinate representation of a given quantum state against the Pauli polytope, the set of all possible occupations according to the Pauli constraints ($0 \le n_i \le 1$), as well as the generalized Pauli polytope, the set of all possible occupations according to the generalized Pauli constraint (Eq. (\ref{eq:paulicons})).  [See Fig. \ref{fig:all_scan_poly}.]
\color{black}




Scatter plots of the occupation numbers for simulated and mitigated, experimental calculations (see Methodology for discussion on error mitigation) are shown in Fig. \ref{fig:all_scan_poly} against the Pauli polytope (the combination of the yellow and blue regions allowed by  $0\le n_i \le 1$) and the generalized Pauli polytope (only the yellow region allowed by Eq. (\ref{eq:paulicons})).  For the simulated calculations (Fig. \ref{fig:sim_0}), possible combinations of angles $\theta_1$, $\theta_2$, and $\theta_3$, varied systematically for $\theta\in\left[0,\frac{\pi}{2}\right]$, are used to prepare quantum states according to Eq. (\ref{eq:prep}).  Note that the darker the color red for a sphere in the figure, the larger the $\lambda_G$ value associated with the given calculation.  As can be seen from Fig. \ref{fig:all_scan_poly}, while the preparations span all orbital occupations consistent with the generalized Pauli constraints and hence all electron correlations, only values approaching the $(n_4,n_5,n_6)=(0.5,0.5,0.5)$ corner of the polytope, known to be the occupations of the GHZ state, demonstrate maximal exciton condensation.  The mitigated, experimental results shown in Fig. \ref{fig:exp_0} in which $\theta_1$ is constrained to either $0$ or $\frac{\pi}{2}$ to limit computational expense confirm the simulation results.

\begin{figure}[tbh!]
  \centering
  \sidesubfloat[Simulation]{\label{fig:sim_0}\includegraphics[scale=.32]{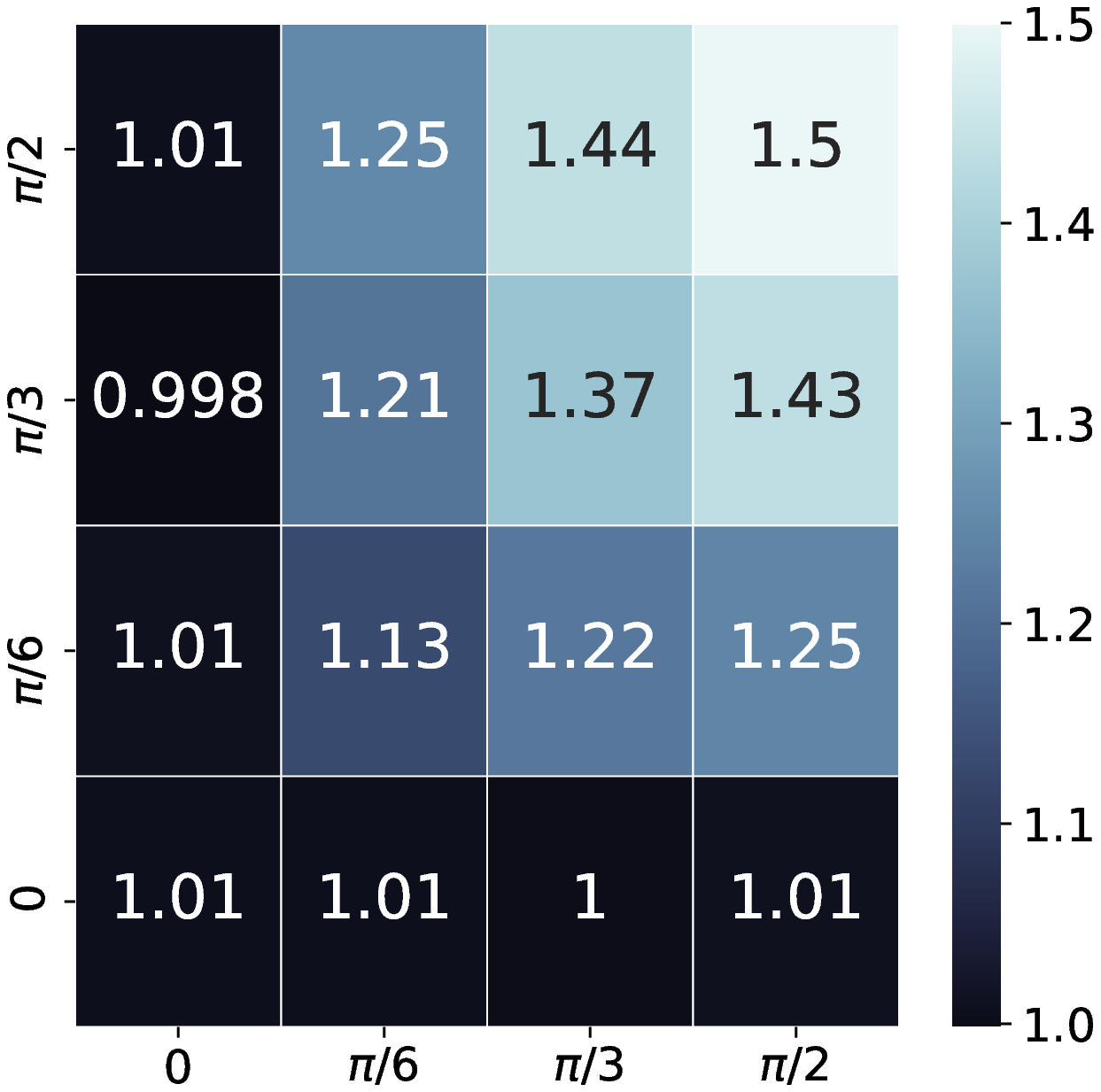}}\\
  \sidesubfloat[Experimental]{\label{fig:exp_0}\includegraphics[scale=.32]{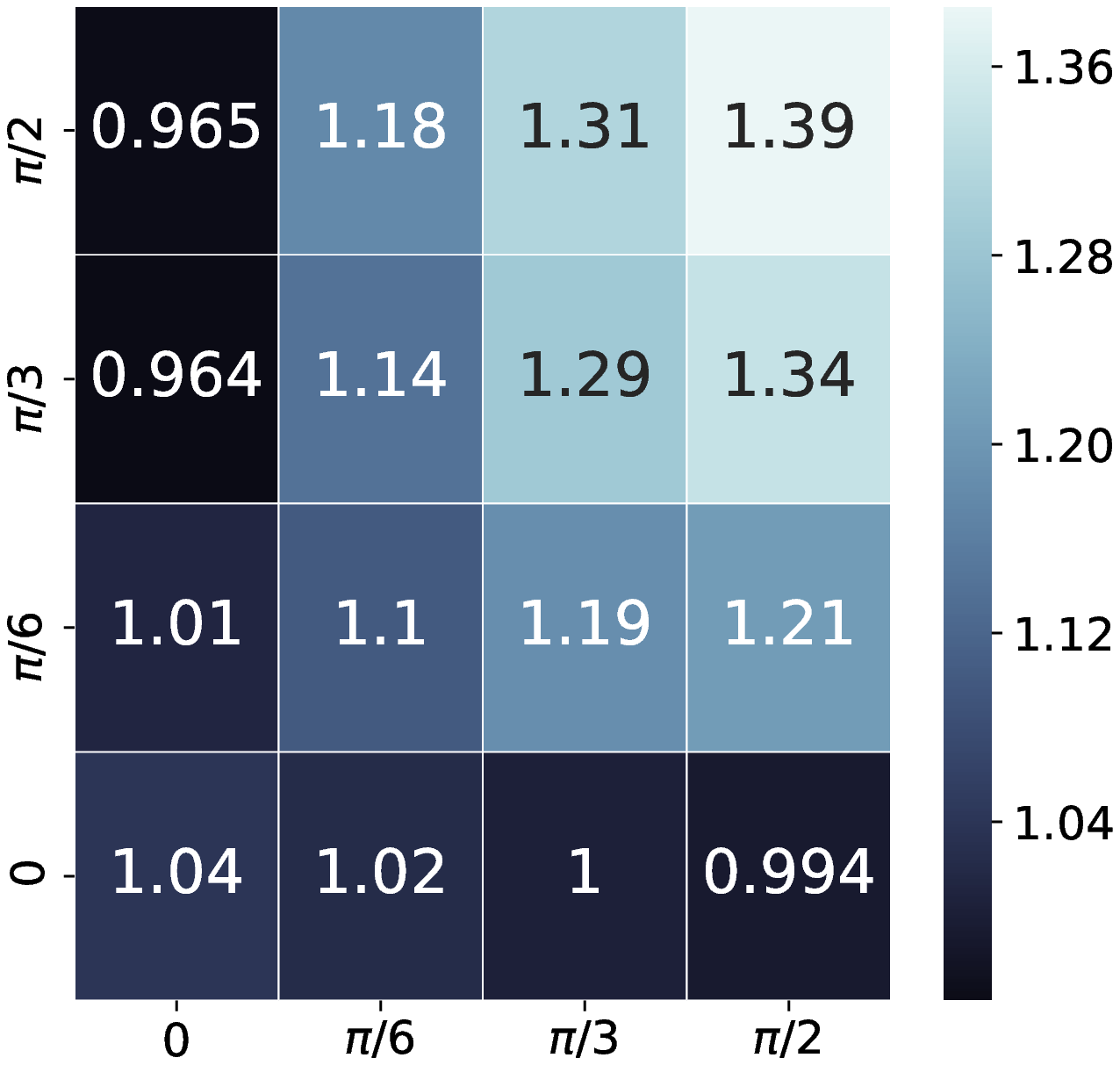}}\\
  \sidesubfloat[Mitigated]{\label{fig:mit_0}\includegraphics[scale=.32]{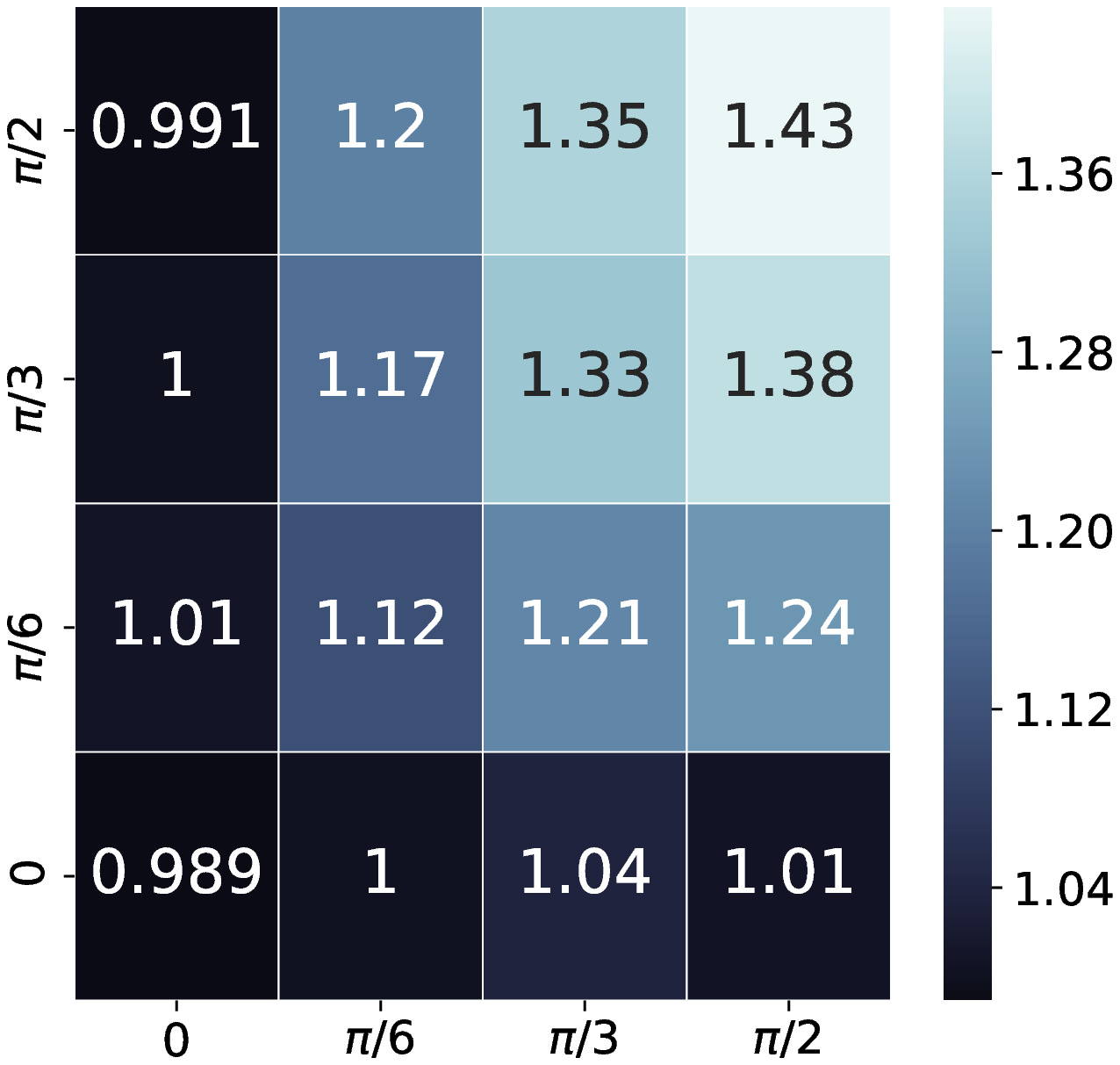}}\\
  \caption{\textbf{Large Particle-hole Eigenvalues and Exciton Condensation.} The largest eigenvalue $\lambda_{G}$ of the modified particle-hole density matrix is shown as a function of the preparation angles $\theta_{2}$ and $\theta_{3}$ in the range $[0, \frac{\pi}{2}]$ with $\theta_{1} = 0$ in Eq. (\ref{eq:prep}) for (a) simulated calculations, (b) experimental results, and (c) mitigated, experimental results.}
\label{fig:0_scan}
\end{figure}
\begin{table}[tbh]
\caption{\color{black} Eigenvalue table for the $^{2}\Tilde{G}$ matrix for simulated ($\lambda_G^{sim.}$), mitigated experimental ($\lambda_G^{mit.}$), and experimental ($\lambda_G^{exp.}$)
GHZ state results}
\begin{tabular}{>{\color{black}}c|>{\color{black}}c|>{\color{black}}c|>{\color{black}}c }
     \textbf{N} & \textbf{$\lambda_G^{sim.}$} & \textbf{$\lambda_G^{mit.}$} & \textbf{$\lambda_G^{exp.}$}\\
    \hline
    3 & 1.50 & 1.44 & 1.39 \\
    4 & 2.00 & 1.92 & 1.80 \\
    5 & 2.50 & 2.33 & 2.22 \\
    6 & 3.00 &  2.70 & 2.27 \\
    7 & 3.50 & ---& 2.44  \\
    8 & 4.00 &  ---  &  2.72  \\
    9 & 4.50 &  ---  &  2.73  \\
    10 & 5.00 &  ---  &  2.93  \\
    11 & 5.50  &  ---&  3.08  \\
    12 &  6.00  &  ---  &  3.22  \\
    13 &  6.50  &  ---  &  3.28  \\
    14 &  7.00  &  ---  &  3.25  \\
    15 &  7.50  &  ---  & \  2.91\footnote{\label{third}\color{black}No suitable circuit orientation on quantum device for creation of a 15 qubit GHZ state, introducing excess error to the calculation.}  \\
    16 &  8.00  & ---  &  2.28  \\
    22 &  11.0  & ---  &  2.68  \\
    28 &  14.0  & ---  &  3.48  \\
    34 &  17.0  & ---  &  3.25  \\
    39 &  19.5  & ---  &  2.71  \\
    47 &  23.5  &  ---  &  2.77  \\
    53 &  26.5  &  ---  &  \ \  3.07\footref{third}
\label{table:all}
\end{tabular}
\end{table}

In order to better visualize the variation of exciton condensate character with respect to variation in the preparation of the qubit quantum state, one particular scan of $\lambda_{G}$ for the minimalistic three-qubit state preparation is shown in Fig. \ref{fig:0_scan} in which the $\theta_1$ value is set to zero and the other angles are varied systematically from $0$ to $\frac{\pi}{2}$ with a $\frac{\pi}{6}$ interval.  In Fig. \ref{fig:0_scan}, results are given for (a) simulation, (b) experiments without mitigation, and (c) mitigated experiments. Note that these particular scanning parameters are chosen as they \color{black} well-represent \color{black} the observed range in $\lambda_G$ and demonstrate the maximal three-qubit $\lambda_G$ of $\frac{N}{2}=\frac{3}{2}=1.5$ for the simulated results.  Additionally, note that even the unmitigated, experimental results [(b)] demonstrate a relatively large $\lambda_G$ of 1.39, a clear demonstration of exciton condensate character despite experimental errors (see Methods for discussion of errors).  This large, non-error-corrected signature of exciton condensation shows that exciton condensation is indeed created on the quantum computer and is not an artifact of error correction.   The large eigenvalue $\lambda_G$ and the degree of saturation of the generalized Pauli constraint in Eq.~(\ref{eq:paulicons}) are reported in Tables~IV through VIII and Fig.~1 for many sets of orbital occupations in the Supporting Information.

\noindent\textbf{Exploration of $\mathbf{\lambda_G}$ for GHZ State.} As shown above, the region of the Pauli polytope associated with the GHZ state, the state described by
\begin{equation}
|\Psi_{GHZ}\rangle=\frac{1}{\sqrt{2}}\left(|0\rangle^{\otimes N}+|1\rangle^{\otimes N}\right)
\label{GHZ_gen}
\end{equation}
for an $N$-qubit system, demonstrates maximal exciton condensate character for three qubits; however, the minimalistic preparation used to probe $\lambda_G$ permits only double excitations, precluding the measurement of $\lambda_G$ associated with the true GHZ state for three qubits.  Therefore, a different qubit preparation scheme is used to  generate the three-qubit GHZ state (see the Methods section).  The maximum exciton condensate character ($\frac{N}{2}=\frac{3}{2}=1.5$) is indeed observed for \color{black} simulation of \color{black} the three-qubit GHZ state.

The GHZ qubit preparation is generalizable to any $N$-qubit state, allowing for the extension of the \color{black} above \color{black} result to larger numbers of qubits, \color{black} the outcomes of which can be seen in Table \ref{table:all}.  \color{black} These results demonstrate that the beginnings of exciton condensation is achieved on quantum computers using $3$-to-$53$ qubits. Note that error mitigation is only feasible for systems with $N \le 6$ qubits as larger-qubit error mitigation schemes necessitate more circuits than the experimental quantum devices allow.  Additionally, to limit computational expense, only real contributions to the reduced density matrices are computed.  See the Methods section for specific experimental details and Tables~II, III, and IV in the Supporting Information for device specifications.
\color{black}

\begin{figure}[tbh!]
  \sidesubfloat[Mitigated]{\label{fig:mitGHZeigs}\includegraphics[width=9cm,angle=0]{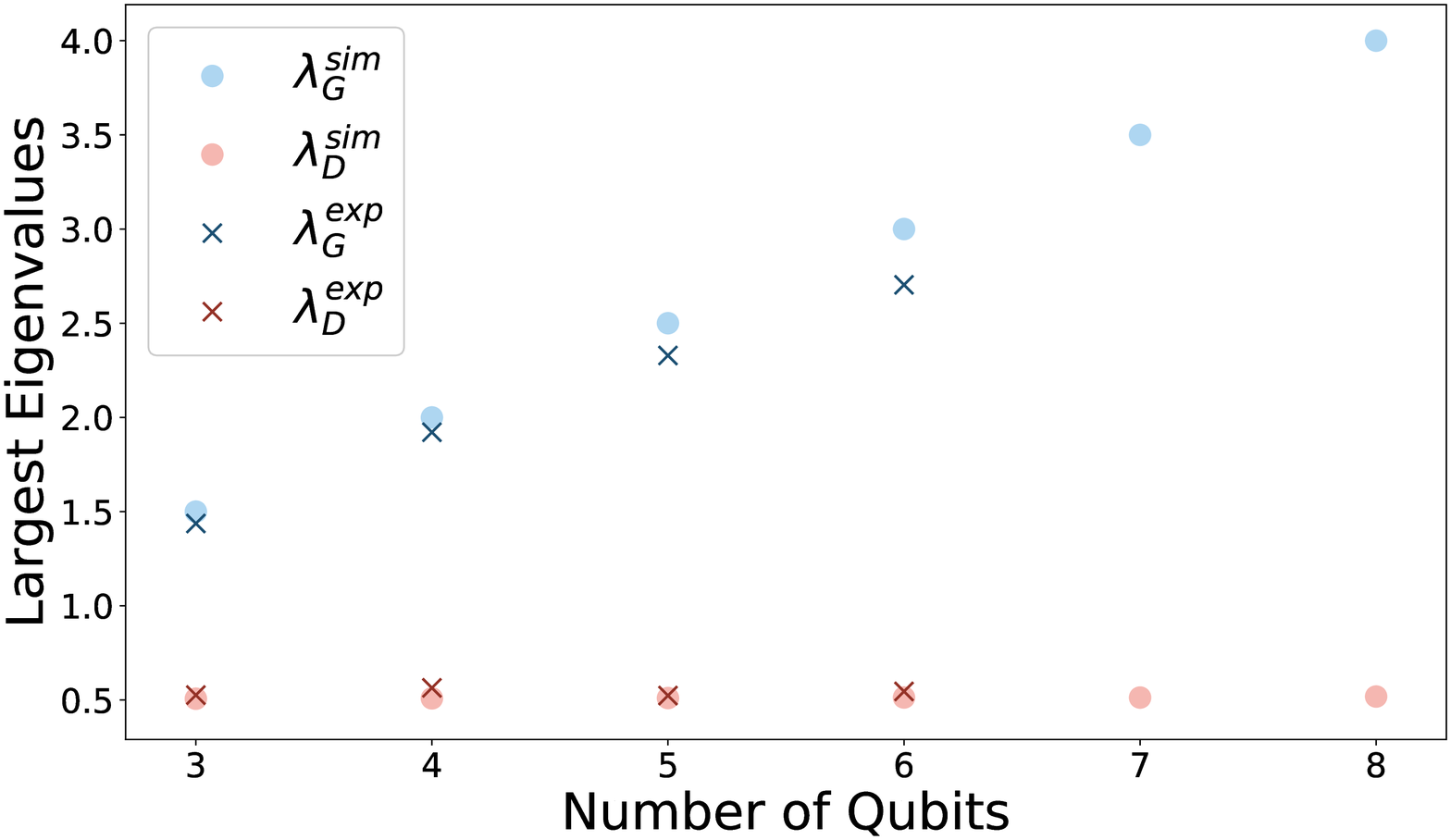}} \\
  \sidesubfloat[All $\lambda_G$]{\label{fig:allGHZeigs}\includegraphics[width=9cm,angle=0]{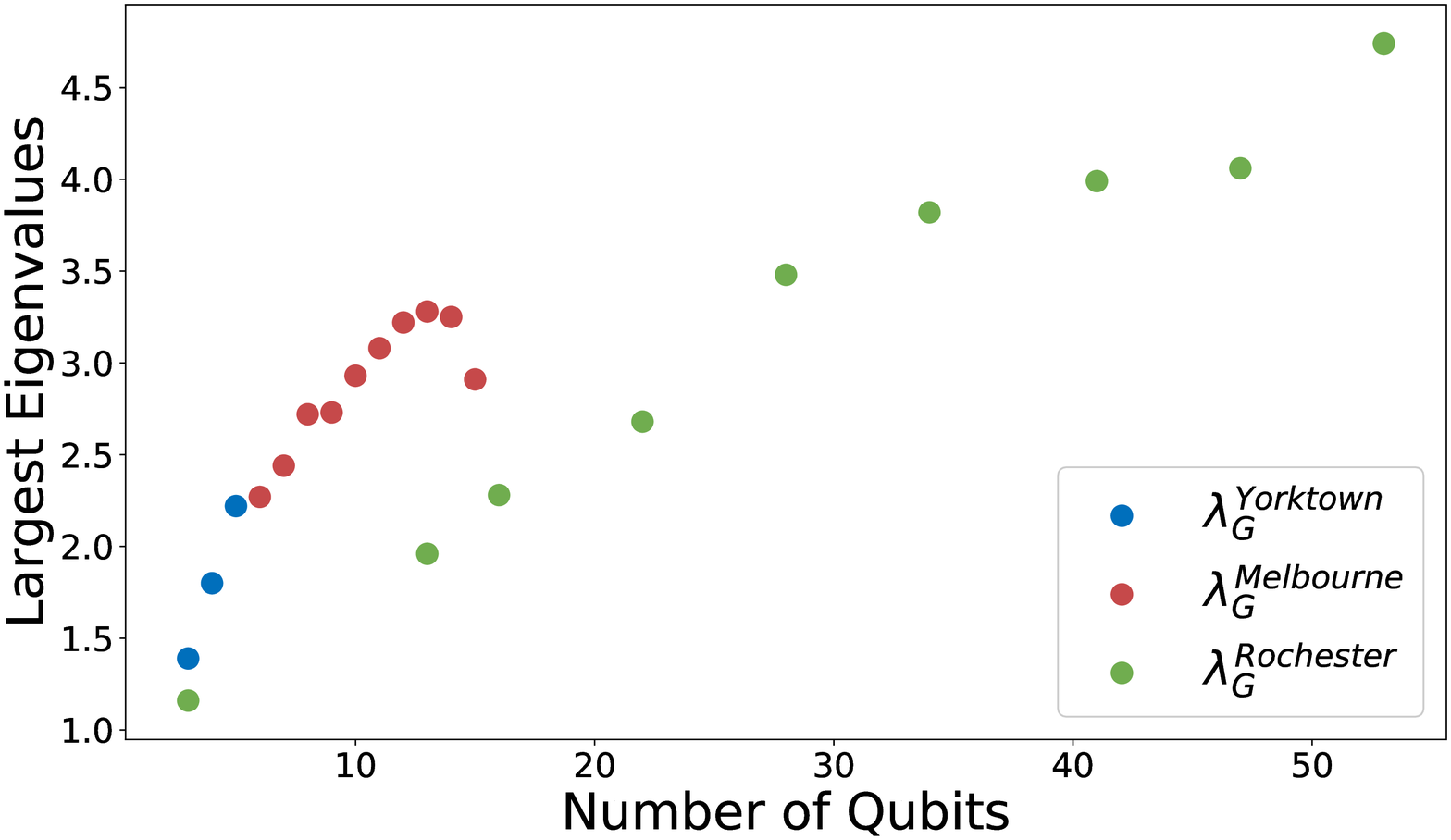}} \\
  \sidesubfloat[Sum of $\lambda$s]{\label{fig:sumGHZeigs}\includegraphics[width=9cm,angle=0]{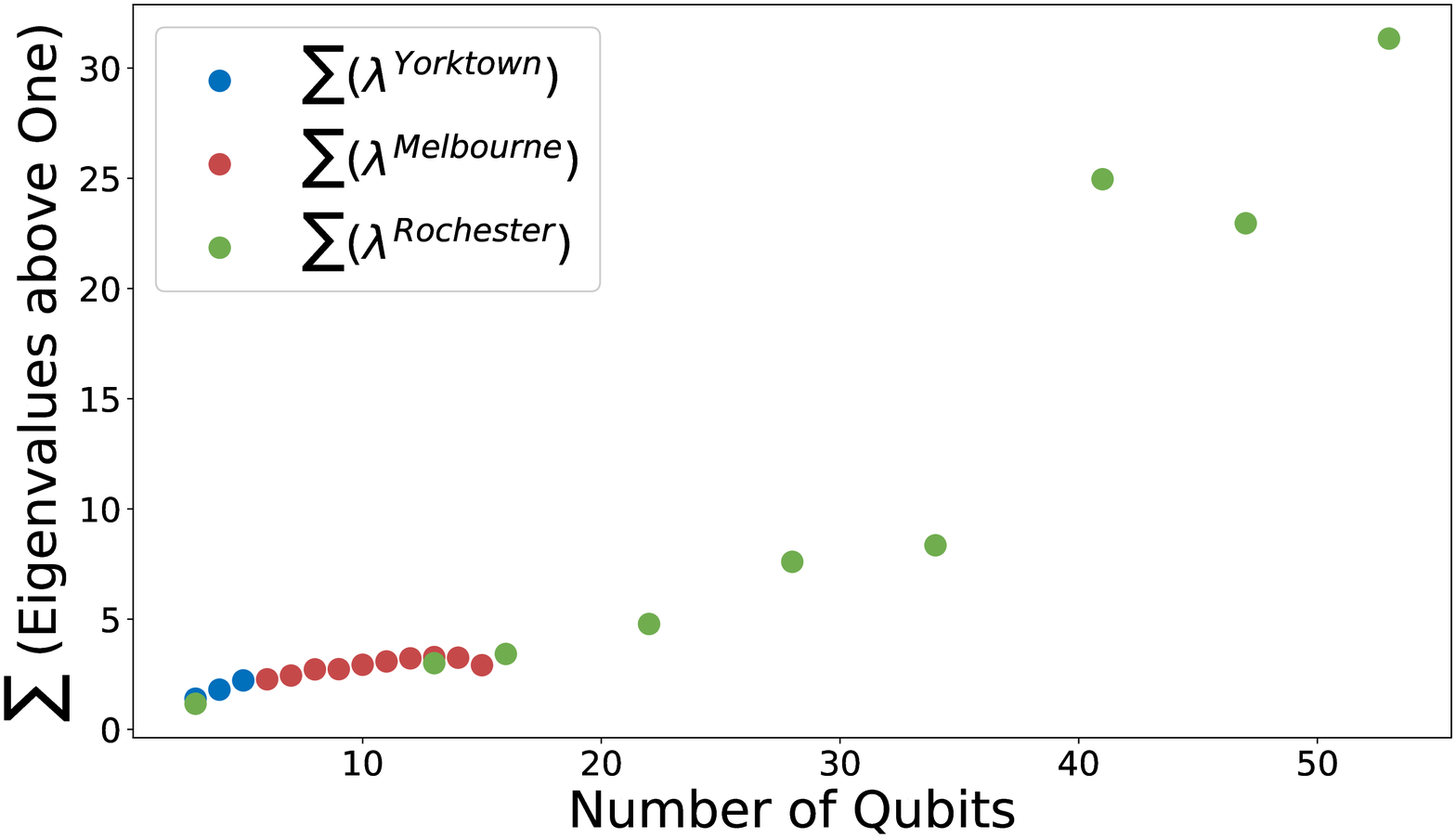}} \\
  \caption{\color{black}\textbf{Exciton Condensation in the GHZ State.} (a) The largest eigenvalue of the ${}^{2}\tilde{G}$ matrix (blue) and the ${}^{2}D$ matrix
(red) are shown for simulated (dots) and all mitigated, experimental ($\times$'s) calculations. (b) The largest eigenvalue of the unmitigated, experimental ${}^{2}\tilde{G}$ matrix  and the (c) sum of all eigenvalues of the unmitigated, experimental ${}^{2}\tilde{G}$ matrix demonstrating exciton condensation ($\lambda>1$) for experiments of $N$ qubits on the Yorktown 5-qubits IBM Quantum Experience device (blue), the Melbourne 15-qubit IBM Quantum Experience device (red), and the Rochester 53-qubit IBM Quantum Experience device (green) are shown.\color{black}}
  \label{fig:GHZeigs}
\end{figure}

As is apparent from simulated results \color{black} (Table \ref{table:all} and Fig. \ref{fig:mitGHZeigs}), \color{black} the GHZ state for all qubits approaches the maximal value for exciton condensate character of $\frac{N}{2}$.  As such, the GHZ state is expected to demonstrate maximal exciton condensation for a given number of qubits.  \color{black} While the experimental results in Table \ref{table:all} and Fig. \ref{fig:allGHZeigs} do not achieve maximal $\lambda_G$ values, although the error-mitigated results do appear to approach $\frac{N}{2}$, exciton condensation character ($\lambda_G > 1$) is observed for each GHZ state prepared for $N=3$-to-$53$ qubits.

\color{black} The larger deviation from the simulated results observed in the higher-qubit experiments\color{black}---in which there seems to be a maximal signature of exciton condensation of around $\lambda_G\approx 3$ (Table \ref{table:all} and Fig. \ref{fig:allGHZeigs})---\color{black} is likely due to the cumulative effects of errors that become increasingly significant as the number of qubits---and hence the number of CNOT gates applied---increases.  (See the Supporting Information for details of gate errors, readout errors, and multi-qubit CNOT errors for the quantum devices employed for experimentation.)  \color{black} These errors seem to prevent the formation of a global excitonic state due to dispersion; however, as the number of qubits is increased, the condensation behavior of the $N$-qubit system does still increase as is shown in Fig. \ref{fig:sumGHZeigs}.  In these higher-qubit experiments, there are multiple eigenvalues of the ${}^{2}\tilde{G}$ matrix above one, indicating that there are multiple eigenstates demonstrating character of exciton condensation.  The sum of the eigenvalues above one increases in an almost linear fashion as the number of qubits is increased, demonstrating an overall increase in the excitonic nature of the prepared states even if maximal condensation behavior in a single orbital can not be obtained for these higher-qubit experiments due to dispersion.

\color{black} The GHZ state is often referred to as a ``the maximally-entangled state'' as it has maximum entanglement entropy \cite{wge_2016}; however, there are diverse types of non-equivalent multi-partite entanglement.  For example, Bose-Einstein condensation, fermion pair condensation, and exciton condensation are all phenomena that occur due to the entanglement of bosons, differing in their signatures and the types of bosons that are entangled.  \color{black} Here we have demonstrated a new characteristic of the maximal entanglement of the GHZ state---namely the maximal entanglement of particle-hole pairs (excitons).  \color{black} Further, the fermion pair condensate character ($\lambda_D$) is additionally probed for the GHZ state, and no fermion pair condensation is observed ($\lambda_D<1$).  (See Fig. \ref{fig:GHZeigs}.)  As such, we have shown that the maximal entanglement of the GHZ state does not correspond to the entanglement of particle-particle pairing. \color{black}

\color{black}

\vspace{0.1cm}
\noindent\textbf{\large{Discussion and Conclusions:}}

\noindent In this study, we have prepared \color{black} molecular-scale exciton condensates for three- to fifty-three-qubit systems on three quantum computers and verified the presence of the condensation through post-measurement computation of the exciton condensate's quantum signature \cite{GR1969}. \color{black} The maximal condensate character is observed for the Greenberg-Horne-Zeilinger (GHZ) ``maximally-entangled'' state, \color{black} indicating that a characteristic of this maximally-entanglement state is the entanglement of particle-hole pairs (excitons). \color{black}  Further, as the transmon qubit quantum states are occupied by microwave photons, the exciton condensate formed by preparation of the GHZ state corresponds to an exciton condensate of photons---the entanglement of photon-hole pairs.  Whether photon-hole pair condensation would have similar properties to those of traditional fermion-hole exciton condensates is unknown, but it seems likely that, \color{black} as the photons are directly analogous to fermions in a traditional exciton condensate, \color{black} the superfluidity of photon-hole excitons should allow for the dissipationless flow of energy, which has possible applications in energy transport.

Additionally, the recognition of the GHZ state as an exciton condensate on a quantum computer establishes a new avenue for the creation and characterization of exciton condensates. As the GHZ state can be remotely and reliably constructed and probed through the use of cloud-accessible quantum devices, this preparation of exciton condensation may be more convenient than prior experimental methodologies. \color{black} Moreover, depending on the type of qubit comprising the quantum device employed for a given experiment, various types of exciton condensates can be prepared. \color{black} As quantum devices are created with a larger number of qubits, preparation of these higher-qubit GHZ states would create more macroscopically-scaled exciton condensates of these various compositions, although---as we have demonstrated---unless sufficient effort is done to prevent dispersion, this condensate character will be scattered throughout multiple eigenstates of the particle-hole RDM.   \color{black} Thus, future exploration of the properties of exciton condensates on quantum computers is anticipated\color{black}, and the creation and characterization of exciton condensates is yet another motivation for the development of low-error quantum devices with macroscopic numbers of qubits. \color{black}


\vspace{0.1cm}
\noindent\textbf{\large{Methods:}}

\noindent We include details on the quantum algorithms used to prepare the qubit states presented in the article; the quantum tomography of the modified particle-hole reduced density matrix; the methodology by which error is mitigated; and relevant details on the experimental quantum \color{black}devices \color{black}
employed.

\noindent\textbf{Quantum algorithms for state preparation.} Two algorithms are utilized in this work to prepare the qubit states.

\textit{Minimalistic, Scanning Approach---}The first algorithm takes a minimalistic approach to span all valid one-qubit occupation numbers for a three qubit system and is given as follows:
\begin{equation}
    |\Psi\rangle=C_3^2R_{y,3}(\theta_3)C_1^2R_{y,1}(\theta_2)C_1^3R_{y,1}(\theta_1)|0\rangle^{\otimes 3},
\label{preparation}
\end{equation}
where $R_{y,i}$ refers to rotation of a qubit $i$ about its Bloch sphere's y-axis---which is given by
\begin{equation}
	R_{y,i}=\left(\begin{array}{cc}
   \cos\left(\frac{\theta}{2}\right) & -\sin\left(\frac{\theta}{2}\right) \\
    \sin\left(\frac{\theta}{2}\right) & \  \  \ \cos\left(\frac{\theta}{2}\right)
\end{array}\right)
\label{eq:rotation}
\end{equation}---and
$C_i^j$ is a standard controlled-NOT (CNOT) gate with control and target qubits $i$ and $j$ respectively.  Note that the control qubit is rotated prior to the application of the CNOT transformation.  Overall, the sequence of transformations in Eq. (\ref{preparation}) yields a wavefunction of the form
\begin{equation}
	|\Psi\rangle=\alpha|000\rangle+\beta|011\rangle+\gamma|101\rangle+\delta|110\rangle,
\label{eq:wavefunc}
\end{equation}
such that $\alpha$, $\beta$, $\gamma$, and $\delta$ are functions of the input angles ($\theta_1$, $\theta_2$, $\theta_3$) and the 1-RDM contains solely diagonal elements.

\textit{The GHZ State---}The GHZ State described in Eq. (\ref{GHZ_gen}) is prepared according to
\begin{equation}
|\Psi\rangle=C_{N-1}^{N}\cdots C_2^3C_1^2H_1|0\rangle^{\otimes N}
\label{GHZ_prep}
\end{equation}
for an $N$-qubit state where $H_i$ represents the Hadamard gate---which maps $|0\rangle$ to $\frac{|0\rangle+|1\rangle}{\sqrt{2}}$ and $|1\rangle$ to $\frac{|0\rangle-|1\rangle}{\sqrt{2}}$---acting on qubit $i$.  There has been much study on the optimal preparation of the GHZ state for various numbers of qubits \cite{cruz_2019}; in this study, however, the simple algorithm from Ref. \onlinecite{TR_GHZ} for GHZ state preparation is utilized as it is easily implemented and generalizable to any arbitrary number of qubits.

\noindent\textbf{Quantum tomography for the modified particle-hole RDM.} The modified particle-hole RDM---with elements given by Eq. (\ref{G2tilde})---is obtained through the translation of all of its elements into the bases of Pauli matrices, which are directly probed on the quantum computer.

First, let us focus on the $^{1}D_i^j$ terms of the ${}^{2}\tilde{G}$ matrix elements.  \color{black} As 1-RDMs simplify to block-diagonal forms with respect to single qubits, there are no non-zero two-qubit 1-RDM terms.  In order for a 1-RDM to be non-zero, then, it must be a one-qubit 1-RDM of the form
\begin{equation}
	\begin{array}{c|cc}
	& {\scriptscriptstyle\hat{a}_{p,0}}& {\scriptscriptstyle\hat{a}_{p,1}} \\\hline
	{\scriptscriptstyle\hat{a}^\dagger_{p,0}} & {\scriptscriptstyle\hat{a}^\dagger_{p,0}\hat{a}_{p,0}} & {\scriptscriptstyle\hat{a}^\dagger_{p,0}\hat{a}_{p,1}} \\
	{\scriptscriptstyle\hat{a}^\dagger_{p,1}} & {\scriptscriptstyle\hat{a}^\dagger_{p,1}\hat{a}_{p,0}} & {\scriptscriptstyle\hat{a}^\dagger_{p,1}\hat{a}_{p,1}}
	\end{array}
\label{form1RDM}
\end{equation}
for qubit $p$  where $\hat{a}_{p}^{\dagger}$ and $\hat{a}_p$ are creation and annihilation operators for qubit $p$, respectively.   \color{black} Each term of these non-zero, one-qubit 1-RDMs \color{black} can be written as a linear combination of Pauli matrices.  For example, $\hat{a}^\dagger_{p,0}\hat{a}_{p,1}$---which represents the qubit going from state $|1\rangle=\left(\begin{array}{c}0 \\ 1\end{array}\right)$ to state $|0\rangle=\left(\begin{array}{c}1 \\ 0\end{array}\right)$---can be written as follows:
\begin{equation}
	\hat{a}^\dagger_{p,0}\hat{a}_{p,1}=\left(\begin{array}{cc}
   0  & 1\\
    0 & 0
\end{array}\right)=\frac{1}{2}\left(X_p+iY_p\right).
\label{eq:translateex}
\end{equation}
Similarly, the other elements can be represented as shown:
\begin{gather}
	\hat{a}^\dagger_{p,1}\hat{a}_{p,0}=
	\left(\begin{array}{cc}
  	 0  & 0 \\
   	 1 & 0
	\end{array}\right)=\frac{1}{2}\left(X_p-iY_p\right), \\
	\hat{a}^\dagger_{p,0}\hat{a}_{p,0}=\left(\begin{array}{cc}
   	1  & 0\\
    	0 & 0
	\end{array}\right)=\frac{1}{2}\left(\hat{I}+Z_p\right), \\
	\hat{a}^\dagger_{p,1}\hat{a}_{p,1}=\left(\begin{array}{cc}
  	 0  & 0 \\
    	0 & 1
	\end{array}\right)=\frac{1}{2}\left(\hat{I}-Z_p\right).
\end{gather}
The expectation value of each matrix element for a given qubit ($p$) can then be obtained by directly probing the expectation values of $X_p$, $Y_p$, and $Z_p$ for a given state preparation.

The overall particle-hole RDM (${}^{2}G$ matrix) can be represented as a $4N \times 4N$ matrix composed of $N^2$ $4 \times 4$ sub-matrices of the form
\begin{equation}
	\begin{array}{c|cccc}
     &  {\scriptscriptstyle\hat{a}_{q,0}^\dagger\hat{a}_{q,0}} &  {\scriptscriptstyle\hat{a}_{q,1}^\dagger\hat{a}_{q,0}} &  {\scriptscriptstyle\hat{a}_{q,0}^\dagger\hat{a}_{q,1}} &  {\scriptscriptstyle\hat{a}_{p,1}^\dagger\hat{a}_{p,1}}\\ \hline
   {\scriptscriptstyle\hat{a}_{p,0}^\dagger\hat{a}_{p,0}} & {\scriptscriptstyle \hat{a}_{p,0}^\dagger\hat{a}_{p,0}\hat{a}_{q,0}^\dagger\hat{a}_{q,0}}  &  {\scriptscriptstyle\hat{a}_{p,0}^\dagger\hat{a}_{p,0}\hat{a}_{q,1}^\dagger\hat{a}_{q,0}} &  {\scriptscriptstyle\hat{a}_{p,0}^\dagger\hat{a}_{p,0}\hat{a}_{q,0}^\dagger\hat{a}_{q,1}} &  {\scriptscriptstyle\hat{a}_{p,0}^\dagger\hat{a}_{p,0}\hat{a}_{p,1}^\dagger\hat{a}_{p,1}}\\
  {\scriptscriptstyle \hat{a}_{p,0}^\dagger\hat{a}_{p,1}} &  {\scriptscriptstyle\hat{a}_{p,0}^\dagger\hat{a}_{p,1}\hat{a}_{q,0}^\dagger\hat{a}_{q,0}}  &  {\scriptscriptstyle\hat{a}_{p,0}^\dagger\hat{a}_{p,1}\hat{a}_{q,1}^\dagger\hat{a}_{q,0}} &  {\scriptscriptstyle\hat{a}_{p,0}^\dagger\hat{a}_{p,1}\hat{a}_{q,0}^\dagger\hat{a}_{q,1}} &  {\scriptscriptstyle\hat{a}_{p,0}^\dagger\hat{a}_{p,1}\hat{a}_{p,1}^\dagger\hat{a}_{p,1}}\\
   {\scriptscriptstyle\hat{a}_{p,1}^\dagger\hat{a}_{p,0}} &  {\scriptscriptstyle\hat{a}_{p,1}^\dagger\hat{a}_{p,0}\hat{a}_{q,0}^\dagger\hat{a}_{q,0}}  &  {\scriptscriptstyle\hat{a}_{p,1}^\dagger\hat{a}_{p,0}\hat{a}_{q,1}^\dagger\hat{a}_{q,0}} &  {\scriptscriptstyle\hat{a}_{p,1}^\dagger\hat{a}_{p,0}\hat{a}_{q,0}^\dagger\hat{a}_{q,1} }& {\scriptscriptstyle \hat{a}_{p,1}^\dagger\hat{a}_{p,0}\hat{a}_{p,1}^\dagger\hat{a}_{p,1}}\\
   {\scriptscriptstyle\hat{a}_{p,1}^\dagger\hat{a}_{p,1}} &  {\scriptscriptstyle\hat{a}_{p,1}^\dagger\hat{a}_{p,1}\hat{a}_{q,0}^\dagger\hat{a}_{q,0}}   &  {\scriptscriptstyle\hat{a}_{p,1}^\dagger\hat{a}_{p,1}\hat{a}_{q,1}^\dagger\hat{a}_{q,0}} &  {\scriptscriptstyle\hat{a}_{p,1}^\dagger\hat{a}_{p,1}\hat{a}_{q,0}^\dagger\hat{a}_{q,1}} &  {\scriptscriptstyle\hat{a}_{p,1}^\dagger\hat{a}_{p,1}\hat{a}_{p,1}^\dagger\hat{a}_{p,1}}.
\end{array}
\label{eq:submat}
\end{equation}
\color{black}where each element of the matrix is the expectation value of the creation and annhilation operator terms shown.\color{black}
  As multi-qubit wavefunctions are the tensor products of individual qubit wavefunctions, these four-body terms can be represented as the tensor products of the two-body terms composing them.  For example, the expectation value of the term $\hat{a}^\dagger_{p,0}\hat{a}_{p,1}\hat{a}^\dagger_{q,1}\hat{a}_{q,1}$ can be written as
\begin{equation}
	\frac{1}{4}\left[\langle X_p\otimes \hat{I}_q\rangle-\langle X_p\otimes Z_q\rangle+i\langle Y_p\otimes \hat{I}_q\rangle-i\langle Y_p\otimes Z_q\rangle \right],
\label{tensex}
\end{equation}
where $\langle Y_p \otimes Z_q\rangle$ is one of nine two-qubit expectation values that can be obtained from quantum computation.  All other terms can be determined using analogous, straightforward methodologies.

Similarly, the overall modified particle-hole RDM (${}^{2}\tilde{G}$ matrix) can be represented as a $4N \times 4N$ matrix composed of $N^2$ $4 \times 4$ sub-matrices.  These sub-matrices are identical to the sub-matrices of the ${}^{2}G$ matrix with the  block modification shown below subtracted off to eliminate the extraneous ground-state-to-ground-state transition:
\begin{equation}
	\begin{array}{c|cccc}
     &  {\scriptscriptstyle\hat{a}_{q,0}^\dagger\hat{a}_{q,0}} &  {\scriptscriptstyle\hat{a}_{q,1}^\dagger\hat{a}_{q,0}} &  {\scriptscriptstyle\hat{a}_{q,0}^\dagger\hat{a}_{q,1}} &  {\scriptscriptstyle\hat{a}_{p,1}^\dagger\hat{a}_{p,1}}\\ \hline
   {\scriptscriptstyle\hat{a}_{p,0}^\dagger\hat{a}_{p,0}} &  {\scriptscriptstyle{}^{1}D_i[0,0]{}^{1}D_j[0,0]} &   {\scriptscriptstyle{}^{1}D_i[0,0]{}^{1}D_j[0,1]} &   {\scriptscriptstyle{}^{1}D_i[0,0]{}^{1}D_j[1,0]} &  {\scriptscriptstyle{}^{1}D_i[0,0]{}^{1}D_j[1,1]} \\
  {\scriptscriptstyle \hat{a}_{p,0}^\dagger\hat{a}_{p,1}} &  {\scriptscriptstyle{}^{1}D_i[0,1]{}^{1}D_j[0,0]} &   {\scriptscriptstyle{}^{1}D_i[0,1]{}^{1}D_j[0,1]} &   {\scriptscriptstyle{}^{1}D_i[0,1]{}^{1}D_j[1,0]} &  {\scriptscriptstyle{}^{1}D_i[0,1]{}^{1}D_j[1,1]}  \\
   {\scriptscriptstyle\hat{a}_{p,1}^\dagger\hat{a}_{p,0}} &  {\scriptscriptstyle{}^{1}D_i[1,0]{}^{1}D_j[0,0]} &   {\scriptscriptstyle{}^{1}D_i[1,0]{}^{1}D_j[0,1]} &   {\scriptscriptstyle{}^{1}D_i[1,0]{}^{1}D_j[1,0]} &  {\scriptscriptstyle{}^{1}D_i[1,0]{}^{1}D_j[1,1]} \\
   {\scriptscriptstyle\hat{a}_{p,1}^\dagger\hat{a}_{p,1}} &  {\scriptscriptstyle{}^{1}D_i[1,1]{}^{1}D_j[0,0]} &   {\scriptscriptstyle{}^{1}D_i[1,1]{}^{1}D_j[0,1]} &   {\scriptscriptstyle{}^{1}D_i[1,1]{}^{1}D_j[1,0]} &  {\scriptscriptstyle{}^{1}D_i[1,1]{}^{1}D_j[1,1]}
\end{array}
\label{eq:modification}
\end{equation}
Note that $^{1}D_i$ is the RDM for qubit $i$ described in Eq. (\ref{form1RDM}) and that $^{1}D_i[a,b]$ is the element of that matrix with matrix coordinates $[a,b]$.

The overall form of the ${}^{2}\tilde{G}$ matrix is hence
\begin{equation}
	\begin{array}{|c|c|c|c|}
	\hline
	{\scriptscriptstyle p=0,q=0} & {\scriptscriptstyle p=0, q=1} & \cdots & {\scriptscriptstyle p=0,q=N-1} \\\hline
	{\scriptscriptstyle p=1,q=0} & {\scriptscriptstyle p=1, q=1} &  \cdots & {\scriptscriptstyle p=1,q=N-1} \\\hline
	\vdots & \vdots & \ddots & \vdots \\\hline
	{\scriptscriptstyle p=N-1,q=0 }& {\scriptscriptstyle p=N-1, q=1} & \cdots &{\scriptscriptstyle  p=N-1,q=N-1} \\\hline
	\end{array}
\label{eq:mat}
\end{equation}
where each $p/q$ combination represents one of the previously-specified blocks, i.e., the difference of the matrices given in Eqs. (\ref{eq:submat}) and (\ref{eq:modification}).  The largest eigenvalue of this overall matrix is the $\lambda_G$ value employed throughout this article.

\color{black}
As the states prepared in this study are real wavefunctions, the imaginary components of the RDMs should be approximately zero within a small range dictated by inherent randomness and by the error of the devices.  Therefore, only the five of the possible nine two-qubit expectation values that correspond to real contributions to the RDMs [$\langle X_p \otimes X_q\rangle$, $\langle Y_p \otimes Y_q\rangle$, $\langle Z_p \otimes Z_q\rangle$, $\langle X_p \otimes Z_q\rangle$, and $\langle Z_p \otimes X_q\rangle$] are non-zero and hence essential for construction of the $^{2}\Tilde{G}$ matrix.
While, for the sake of completeness, the negligibly-small imaginary components are included in the construction of the $^{2}\Tilde{G}$ matrix for the low-qubit ($N=3-5$) computations, only real components are included in the $^{2}\Tilde{G}$ matrix for higher-qubit ($N=6-15$) computations to lower computational expense.
\color{black}

\noindent\textbf{Error mitigation.} A measurement correction fitter for a tensored calibration is employed to mitigate measurement error through use of the ``least\_squares''  method---that constrains the resultant mitigated counts to having physical probabilities---to construct a mitigation filter that can be applied to experimental data.

\noindent\textbf{Experimental quantum device specifications.} \color{black} Throughout this work, we  \color{black} employ \color{black} the ibmqx2 \color{black} (ibmq\_5\_yorktown) \color{black}  \cite{qx2_2019}\color{black}, \color{black} the ibmq\_16\_melbourne \cite{melbourne}\color{black}, and the ibmq\_rochester \cite{rochester} \color{black} IBM Quantum Experience devices, which are available online. \color{black} Unless explicitly stated otherwise, \color{black}all low-qubit ($N \le 5$) experiments are conducted using the five-qubit ibmqx2  \color{black} (ibmq\_5\_yorktown) \color{black} device, \color{black} all midrange-qubit ($5<N\le15$) experiments \color{black} are conducted using the fifteen-qubit ibmq\_16\_melbourne device\color{black}, and all high-qubit ($N>15$) experiments are conducted using the fifty-three-qubit ibmq\_rochester device. \color{black} These quantum devices are \color{black} composed of fixed-frequency transmon qubits with co-planer waveguide resonators \cite{koch_2007,chow_2011}.  Experimental calibration data and connectivity for \color{black} these devices are included in the Supporting Information. \color{black}


\vspace{0.2cm}

\newpage
\begin{acknowledgments}
\textit{Acknowledgments}: \color{black} The authors acknowledge use of the IBM Q for this work.
\color{black} D.A.M.   gratefully   acknowledges the  Department of Energy, Office of Basic Energy Sciences, Grant DE-SC0019215, the   U.S.   National Science  Foundation  Grant No.  CHE-1565638
and  the  U.S.  Army Research Office (ARO) Grant No. W911NF-16-1-0152. \color{black} The views expressed are of the authors and do not reflect the official policy or position of IBM or the IBM Q team. \color{black}
\end{acknowledgments}

\noindent {\bf Author contributions.} D. A. M. conceived of the project.  L.M.S. performed the computations on the quantum computers.  L.M.S., S.E.S., and D.A.M. analyzed the data and wrote the manuscript.

\noindent {\bf Competing interests.} The authors do not have competing interests.

\noindent {\bf Materials \& Correspondence.} D. A. M. is the corresponding author at damazz@uchicago.edu.

	\bibliography{references}

\begin{thebibliography}{42}%
\makeatletter
\providecommand \@ifxundefined [1]{%
 \@ifx{#1\undefined}
}%
\providecommand \@ifnum [1]{%
 \ifnum #1\expandafter \@firstoftwo
 \else \expandafter \@secondoftwo
 \fi
}%
\providecommand \@ifx [1]{%
 \ifx #1\expandafter \@firstoftwo
 \else \expandafter \@secondoftwo
 \fi
}%
\providecommand \natexlab [1]{#1}%
\providecommand \enquote  [1]{``#1''}%
\providecommand \bibnamefont  [1]{#1}%
\providecommand \bibfnamefont [1]{#1}%
\providecommand \citenamefont [1]{#1}%
\providecommand \href@noop [0]{\@secondoftwo}%
\providecommand \href [0]{\begingroup \@sanitize@url \@href}%
\providecommand \@href[1]{\@@startlink{#1}\@@href}%
\providecommand \@@href[1]{\endgroup#1\@@endlink}%
\providecommand \@sanitize@url [0]{\catcode `\\12\catcode `\$12\catcode
  `\&12\catcode `\#12\catcode `\^12\catcode `\_12\catcode `\%12\relax}%
\providecommand \@@startlink[1]{}%
\providecommand \@@endlink[0]{}%
\providecommand \url  [0]{\begingroup\@sanitize@url \@url }%
\providecommand \@url [1]{\endgroup\@href {#1}{\urlprefix }}%
\providecommand \urlprefix  [0]{URL }%
\providecommand \Eprint [0]{\href }%
\providecommand \doibase [0]{http://dx.doi.org/}%
\providecommand \selectlanguage [0]{\@gobble}%
\providecommand \bibinfo  [0]{\@secondoftwo}%
\providecommand \bibfield  [0]{\@secondoftwo}%
\providecommand \translation [1]{[#1]}%
\providecommand \BibitemOpen [0]{}%
\providecommand \bibitemStop [0]{}%
\providecommand \bibitemNoStop [0]{.\EOS\space}%
\providecommand \EOS [0]{\spacefactor3000\relax}%
\providecommand \BibitemShut  [1]{\csname bibitem#1\endcsname}%
\let\auto@bib@innerbib\@empty
\bibitem [{\citenamefont {Fil}\ and\ \citenamefont
  {Shevchenko}(2018)}]{Fil_Shevchenko_Rev}%
  \BibitemOpen
  \bibfield  {author} {\bibinfo {author} {\bibfnamefont {D.~V.}\ \bibnamefont
  {Fil}}\ and\ \bibinfo {author} {\bibfnamefont {S.~I.}\ \bibnamefont
  {Shevchenko}},\ }\bibfield  {title} {\enquote {\bibinfo {title}
  {Electron-hole superconductivity (review)},}\ }\href
  {https://aip.scitation.org/doi/pdf/10.1063/1.5052674?class=pdf} {\bibfield
  {journal} {\bibinfo  {journal} {Low Temp. Phys.}\ }\textbf {\bibinfo {volume}
  {44}},\ \bibinfo {pages} {867--909} (\bibinfo {year} {2018})}\BibitemShut
  {NoStop}%
\bibitem [{\citenamefont {Keldysh}(2017)}]{keldysh_2017}%
  \BibitemOpen
  \bibfield  {author} {\bibinfo {author} {\bibfnamefont {L.~V.}\ \bibnamefont
  {Keldysh}},\ }\bibfield  {title} {\enquote {\bibinfo {title} {Coherent states
  of excitons},}\ }\href {\doibase 10.3367/ufne.2017.10.038227} {\bibfield
  {journal} {\bibinfo  {journal} {Physics-Uspekhi}\ }\textbf {\bibinfo {volume}
  {60}},\ \bibinfo {pages} {1180–1186} (\bibinfo {year} {2017})}\BibitemShut
  {NoStop}%
\bibitem [{\citenamefont {Kellogg}\ \emph {et~al.}(2004)\citenamefont
  {Kellogg}, \citenamefont {Eisenstein}, \citenamefont {Pfeiffer},\ and\
  \citenamefont {West}}]{KSE2004}%
  \BibitemOpen
  \bibfield  {author} {\bibinfo {author} {\bibfnamefont {M.}~\bibnamefont
  {Kellogg}}, \bibinfo {author} {\bibfnamefont {J.~P.}\ \bibnamefont
  {Eisenstein}}, \bibinfo {author} {\bibfnamefont {L.~N.}\ \bibnamefont
  {Pfeiffer}}, \ and\ \bibinfo {author} {\bibfnamefont {K.~W.}\ \bibnamefont
  {West}},\ }\bibfield  {title} {\enquote {\bibinfo {title} {Vanishing {H}all
  resistance at high magnetic field in a double-layer two-dimensional electron
  system},}\ }\href
  {https://journals.aps.org/prl/pdf/10.1103/PhysRevLett.93.036801} {\bibfield
  {journal} {\bibinfo  {journal} {Phys. Rev. Lett.}\ }\textbf {\bibinfo
  {volume} {93}},\ \bibinfo {pages} {036801} (\bibinfo {year}
  {2004})}\BibitemShut {NoStop}%
\bibitem [{\citenamefont {Tutuc}\ \emph {et~al.}(2004)\citenamefont {Tutuc},
  \citenamefont {Shayegan},\ and\ \citenamefont {Huse}}]{TSH2004}%
  \BibitemOpen
  \bibfield  {author} {\bibinfo {author} {\bibfnamefont {E.}~\bibnamefont
  {Tutuc}}, \bibinfo {author} {\bibfnamefont {M.}~\bibnamefont {Shayegan}}, \
  and\ \bibinfo {author} {\bibfnamefont {D.~A.}\ \bibnamefont {Huse}},\
  }\bibfield  {title} {\enquote {\bibinfo {title} {Counterflow measurements in
  strongly correlated {G}a{A}s hole bilayers: evidence for electron-hole
  pairing},}\ }\href {\doibase 10.1103/PhysRevLett.93.036802} {\bibfield
  {journal} {\bibinfo  {journal} {Phys. Rev. Lett.}\ }\textbf {\bibinfo
  {volume} {93}},\ \bibinfo {pages} {36802} (\bibinfo {year}
  {2004})}\BibitemShut {NoStop}%
\bibitem [{\citenamefont {Safaei}\ and\ \citenamefont
  {Mazziotti}(2018)}]{Shiva}%
  \BibitemOpen
  \bibfield  {author} {\bibinfo {author} {\bibfnamefont {S.}~\bibnamefont
  {Safaei}}\ and\ \bibinfo {author} {\bibfnamefont {D.~A.}\ \bibnamefont
  {Mazziotti}},\ }\bibfield  {title} {\enquote {\bibinfo {title} {Quantum
  signature of exciton condensation},}\ }\href
  {https://journals.aps.org/prb/pdf/10.1103/PhysRevB.98.045122/Users/lmsager/Downloads/IOPEXPORT_BIB.bib}
  {\bibfield  {journal} {\bibinfo  {journal} {Phys. Rev. B}\ }\textbf {\bibinfo
  {volume} {98}},\ \bibinfo {pages} {045122} (\bibinfo {year}
  {2018})}\BibitemShut {NoStop}%
\bibitem [{\citenamefont {Kogar}\ \emph {et~al.}(2017)\citenamefont {Kogar},
  \citenamefont {Rak}, \citenamefont {Vig}, \citenamefont {Husain},
  \citenamefont {Flicker}, \citenamefont {Joe}, \citenamefont {Venema},
  \citenamefont {MacDougall}, \citenamefont {Chiang}, \citenamefont {Fradkin},
  \citenamefont {van Wezel},\ and\ \citenamefont {Abbamonte}}]{Kogar2017}%
  \BibitemOpen
  \bibfield  {author} {\bibinfo {author} {\bibfnamefont {A.}~\bibnamefont
  {Kogar}}, \bibinfo {author} {\bibfnamefont {M.~S.}\ \bibnamefont {Rak}},
  \bibinfo {author} {\bibfnamefont {S.}~\bibnamefont {Vig}}, \bibinfo {author}
  {\bibfnamefont {A.~A.}\ \bibnamefont {Husain}}, \bibinfo {author}
  {\bibfnamefont {F.}~\bibnamefont {Flicker}}, \bibinfo {author} {\bibfnamefont
  {Y.~I.}\ \bibnamefont {Joe}}, \bibinfo {author} {\bibfnamefont
  {L.}~\bibnamefont {Venema}}, \bibinfo {author} {\bibfnamefont {G.~J.}\
  \bibnamefont {MacDougall}}, \bibinfo {author} {\bibfnamefont {T.~C.}\
  \bibnamefont {Chiang}}, \bibinfo {author} {\bibfnamefont {E.}~\bibnamefont
  {Fradkin}}, \bibinfo {author} {\bibfnamefont {J.}~\bibnamefont {van Wezel}},
  \ and\ \bibinfo {author} {\bibfnamefont {P.}~\bibnamefont {Abbamonte}},\
  }\bibfield  {title} {\enquote {\bibinfo {title} {Signatures of exciton
  condensation in a transition metal dichalcogenide},}\ }\href {\doibase
  10.1126/science.aam6432} {\bibfield  {journal} {\bibinfo  {journal}
  {Science}\ }\textbf {\bibinfo {volume} {358}},\ \bibinfo {pages} {1314--1317}
  (\bibinfo {year} {2017})}\BibitemShut {NoStop}%
\bibitem [{\citenamefont {Liu}\ \emph {et~al.}(2017)\citenamefont {Liu},
  \citenamefont {Watanabe}, \citenamefont {Taniguchi}, \citenamefont
  {Halperin},\ and\ \citenamefont {Kim}}]{LWT2017}%
  \BibitemOpen
  \bibfield  {author} {\bibinfo {author} {\bibfnamefont {X.}~\bibnamefont
  {Liu}}, \bibinfo {author} {\bibfnamefont {K.}~\bibnamefont {Watanabe}},
  \bibinfo {author} {\bibfnamefont {T.}~\bibnamefont {Taniguchi}}, \bibinfo
  {author} {\bibfnamefont {B.~I.}\ \bibnamefont {Halperin}}, \ and\ \bibinfo
  {author} {\bibfnamefont {P.}~\bibnamefont {Kim}},\ }\bibfield  {title}
  {\enquote {\bibinfo {title} {Quantum {H}all drag of exciton condensate in
  graphene},}\ }\href {\doibase 10.1038/nphys4116} {\bibfield  {journal}
  {\bibinfo  {journal} {Nat. Phys.}\ }\textbf {\bibinfo {volume} {13}},\
  \bibinfo {pages} {746--750} (\bibinfo {year} {2017})}\BibitemShut {NoStop}%
\bibitem [{\citenamefont {Varsano}\ \emph {et~al.}(2017)\citenamefont
  {Varsano}, \citenamefont {Sorella}, \citenamefont {Sangalli}, \citenamefont
  {Barborini}, \citenamefont {Corni}, \citenamefont {Molinari},\ and\
  \citenamefont {Rontani}}]{varsano_2017}%
  \BibitemOpen
  \bibfield  {author} {\bibinfo {author} {\bibfnamefont {D.}~\bibnamefont
  {Varsano}}, \bibinfo {author} {\bibfnamefont {S.}~\bibnamefont {Sorella}},
  \bibinfo {author} {\bibfnamefont {D.}~\bibnamefont {Sangalli}}, \bibinfo
  {author} {\bibfnamefont {M.}~\bibnamefont {Barborini}}, \bibinfo {author}
  {\bibfnamefont {S.}~\bibnamefont {Corni}}, \bibinfo {author} {\bibfnamefont
  {E.}~\bibnamefont {Molinari}}, \ and\ \bibinfo {author} {\bibfnamefont
  {M.}~\bibnamefont {Rontani}},\ }\bibfield  {title} {\enquote {\bibinfo
  {title} {Carbon nanotubes as excitonic insulators},}\ }\href
  {https://www.nature.com/articles/s41467-017-01660-8} {\bibfield  {journal}
  {\bibinfo  {journal} {Nature Communications}\ }\textbf {\bibinfo {volume}
  {8}} (\bibinfo {year} {2017})}\BibitemShut {NoStop}%
\bibitem [{\citenamefont {Kasprzak}\ \emph {et~al.}(2006)\citenamefont
  {Kasprzak}, \citenamefont {Richard}, \citenamefont {Kundermann},
  \citenamefont {Baas}, \citenamefont {Jeambrun}, \citenamefont {Keeling},
  \citenamefont {Marchetti}, \citenamefont {Szymanska}, \citenamefont
  {Andr{\'e}}, \citenamefont {Staehli}, \citenamefont {Savona}, \citenamefont
  {Littlewood}, \citenamefont {Deveaud},\ and\ \citenamefont {Dang}}]{KRK2006}%
  \BibitemOpen
  \bibfield  {author} {\bibinfo {author} {\bibfnamefont {J.}~\bibnamefont
  {Kasprzak}}, \bibinfo {author} {\bibfnamefont {M.}~\bibnamefont {Richard}},
  \bibinfo {author} {\bibfnamefont {S.}~\bibnamefont {Kundermann}}, \bibinfo
  {author} {\bibfnamefont {A.}~\bibnamefont {Baas}}, \bibinfo {author}
  {\bibfnamefont {P.}~\bibnamefont {Jeambrun}}, \bibinfo {author}
  {\bibfnamefont {J.~M.~J.}\ \bibnamefont {Keeling}}, \bibinfo {author}
  {\bibfnamefont {F.~M.}\ \bibnamefont {Marchetti}}, \bibinfo {author}
  {\bibfnamefont {M.~H.}\ \bibnamefont {Szymanska}}, \bibinfo {author}
  {\bibfnamefont {R.}~\bibnamefont {Andr{\'e}}}, \bibinfo {author}
  {\bibfnamefont {J.~L.}\ \bibnamefont {Staehli}}, \bibinfo {author}
  {\bibfnamefont {V.}~\bibnamefont {Savona}}, \bibinfo {author} {\bibfnamefont
  {P.~B.}\ \bibnamefont {Littlewood}}, \bibinfo {author} {\bibfnamefont
  {B.}~\bibnamefont {Deveaud}}, \ and\ \bibinfo {author} {\bibfnamefont
  {Le~Si}\ \bibnamefont {Dang}},\ }\bibfield  {title} {\enquote {\bibinfo
  {title} {{B}ose-{E}instein condensation of exciton polaritons},}\ }\href
  {\doibase 10.1038/nature05131} {\bibfield  {journal} {\bibinfo  {journal}
  {Nature}\ }\textbf {\bibinfo {volume} {443}},\ \bibinfo {pages} {409}
  (\bibinfo {year} {2006})}\BibitemShut {NoStop}%
\bibitem [{\citenamefont {Fuhrer}\ and\ \citenamefont
  {Hamilton}(2016)}]{fuhrer_hamilton_2016}%
  \BibitemOpen
  \bibfield  {author} {\bibinfo {author} {\bibfnamefont {M.~S.}\ \bibnamefont
  {Fuhrer}}\ and\ \bibinfo {author} {\bibfnamefont {A.~R.}\ \bibnamefont
  {Hamilton}},\ }\bibfield  {title} {\enquote {\bibinfo {title} {Chasing the
  exciton condensate},}\ }\href {https://physics.aps.org/articles/v9/80}
  {\bibfield  {journal} {\bibinfo  {journal} {Physics}\ }\textbf {\bibinfo
  {volume} {9}} (\bibinfo {year} {2016})}\BibitemShut {NoStop}%
\bibitem [{\citenamefont {Spielman}\ \emph {et~al.}(2000)\citenamefont
  {Spielman}, \citenamefont {Eisenstein}, \citenamefont {Pfeiffer},\ and\
  \citenamefont {West}}]{SEP2000}%
  \BibitemOpen
  \bibfield  {author} {\bibinfo {author} {\bibfnamefont {I.~B.}\ \bibnamefont
  {Spielman}}, \bibinfo {author} {\bibfnamefont {J.~P.}\ \bibnamefont
  {Eisenstein}}, \bibinfo {author} {\bibfnamefont {L.~N.}\ \bibnamefont
  {Pfeiffer}}, \ and\ \bibinfo {author} {\bibfnamefont {K.~W.}\ \bibnamefont
  {West}},\ }\bibfield  {title} {\enquote {\bibinfo {title} {Resonantly
  enhanced tunneling in a double layer quantum {H}all ferromagnet},}\ }\href
  {\doibase 10.1103/PhysRevLett.84.5808} {\bibfield  {journal} {\bibinfo
  {journal} {Phys. Rev. Lett.}\ }\textbf {\bibinfo {volume} {84}},\ \bibinfo
  {pages} {5808--5811} (\bibinfo {year} {2000})}\BibitemShut {NoStop}%
\bibitem [{\citenamefont {Nandi}\ \emph {et~al.}(2012)\citenamefont {Nandi},
  \citenamefont {Finck}, \citenamefont {Eisenstein}, \citenamefont {Pfeiffer},\
  and\ \citenamefont {West}}]{NFE2012}%
  \BibitemOpen
  \bibfield  {author} {\bibinfo {author} {\bibfnamefont {D.}~\bibnamefont
  {Nandi}}, \bibinfo {author} {\bibfnamefont {A.~D.~K.}\ \bibnamefont {Finck}},
  \bibinfo {author} {\bibfnamefont {J.~P.}\ \bibnamefont {Eisenstein}},
  \bibinfo {author} {\bibfnamefont {L.~N.}\ \bibnamefont {Pfeiffer}}, \ and\
  \bibinfo {author} {\bibfnamefont {K.~W.}\ \bibnamefont {West}},\ }\bibfield
  {title} {\enquote {\bibinfo {title} {Exciton condensation and perfect
  {C}oulomb drag},}\ }\href {\doibase 10.1038/nature11302} {\bibfield
  {journal} {\bibinfo  {journal} {Nature}\ }\textbf {\bibinfo {volume} {488}},\
  \bibinfo {pages} {481--484} (\bibinfo {year} {2012})}\BibitemShut {NoStop}%
\bibitem [{\citenamefont {Ma}\ \emph {et~al.}(2019)\citenamefont {Ma},
  \citenamefont {Saxberg}, \citenamefont {Owens}, \citenamefont {Leung},
  \citenamefont {Lu}, \citenamefont {Simon},\ and\ \citenamefont
  {Schuster}}]{schuster_2019}%
  \BibitemOpen
  \bibfield  {author} {\bibinfo {author} {\bibfnamefont {R.}~\bibnamefont
  {Ma}}, \bibinfo {author} {\bibfnamefont {B.}~\bibnamefont {Saxberg}},
  \bibinfo {author} {\bibfnamefont {C.}~\bibnamefont {Owens}}, \bibinfo
  {author} {\bibfnamefont {N.}~\bibnamefont {Leung}}, \bibinfo {author}
  {\bibfnamefont {Y.}~\bibnamefont {Lu}}, \bibinfo {author} {\bibfnamefont
  {J.}~\bibnamefont {Simon}}, \ and\ \bibinfo {author} {\bibfnamefont {D.~I.}\
  \bibnamefont {Schuster}},\ }\bibfield  {title} {\enquote {\bibinfo {title} {A
  dissipatively stabilized {M}ott insulator of photons},}\ }\href {\doibase
  10.1038/s41586-019-0897-9} {\bibfield  {journal} {\bibinfo  {journal}
  {Nature}\ }\textbf {\bibinfo {volume} {566}},\ \bibinfo {pages} {51–57}
  (\bibinfo {year} {2019})}\BibitemShut {NoStop}%
\bibitem [{\citenamefont {{Lipkin}}\ \emph {et~al.}(1965)\citenamefont
  {{Lipkin}}, \citenamefont {{Meshkov}},\ and\ \citenamefont
  {{Glick}}}]{Lipkin_model}%
  \BibitemOpen
  \bibfield  {author} {\bibinfo {author} {\bibfnamefont {H.~J.}\ \bibnamefont
  {{Lipkin}}}, \bibinfo {author} {\bibfnamefont {N.}~\bibnamefont {{Meshkov}}},
  \ and\ \bibinfo {author} {\bibfnamefont {A.~J.}\ \bibnamefont {{Glick}}},\
  }\bibfield  {title} {\enquote {\bibinfo {title} {Validity of many-body
  approximation methods for a solvable model: (i.) exact solutions and
  perturbation theory},}\ }\href {\doibase 10.1016/0029-5582(65)90862-X}
  {\bibfield  {journal} {\bibinfo  {journal} {Nucl. Phys. A}\ }\textbf
  {\bibinfo {volume} {62}},\ \bibinfo {pages} {188--198} (\bibinfo {year}
  {1965})}\BibitemShut {NoStop}%
\bibitem [{\citenamefont {P\'erez}\ \emph {et~al.}(1988)\citenamefont
  {P\'erez}, \citenamefont {Cambiaggio},\ and\ \citenamefont
  {Vary}}]{texpansion}%
  \BibitemOpen
  \bibfield  {author} {\bibinfo {author} {\bibfnamefont {R.}~\bibnamefont
  {P\'erez}}, \bibinfo {author} {\bibfnamefont {M.~C.}\ \bibnamefont
  {Cambiaggio}}, \ and\ \bibinfo {author} {\bibfnamefont {J.~P.}\ \bibnamefont
  {Vary}},\ }\bibfield  {title} {\enquote {\bibinfo {title} {t expansion and
  the {L}ipkin model},}\ }\href {\doibase 10.1103/PhysRevC.37.2194} {\bibfield
  {journal} {\bibinfo  {journal} {Phys. Rev. C}\ }\textbf {\bibinfo {volume}
  {37}},\ \bibinfo {pages} {2194--2198} (\bibinfo {year} {1988})}\BibitemShut
  {NoStop}%
\bibitem [{\citenamefont {Mazziotti}(1998)}]{David1998}%
  \BibitemOpen
  \bibfield  {author} {\bibinfo {author} {\bibfnamefont {D.~A.}\ \bibnamefont
  {Mazziotti}},\ }\bibfield  {title} {\enquote {\bibinfo {title} {Contracted
  {S}chrödinger equation: determining quantum energies and two-particle
  density matrices without wave functions},}\ }\href {\doibase
  10.1103/PhysRevA.69.012507} {\bibfield  {journal} {\bibinfo  {journal} {Phys.
  Rev. A}\ }\textbf {\bibinfo {volume} {57}},\ \bibinfo {pages} {4219--4234}
  (\bibinfo {year} {1998})}\BibitemShut {NoStop}%
\bibitem [{\citenamefont {Stein}(2000)}]{Stein_2000}%
  \BibitemOpen
  \bibfield  {author} {\bibinfo {author} {\bibfnamefont {J.}~\bibnamefont
  {Stein}},\ }\bibfield  {title} {\enquote {\bibinfo {title} {Unitary flow of
  the bosonized large-{NLipkin} model},}\ }\href {\doibase
  10.1088/0954-3899/26/4/304} {\bibfield  {journal} {\bibinfo  {journal} {J.
  Phys. G Nucl. Part. Phys.}\ }\textbf {\bibinfo {volume} {26}},\ \bibinfo
  {pages} {377--385} (\bibinfo {year} {2000})}\BibitemShut {NoStop}%
\bibitem [{\citenamefont {Mazziotti}(2004)}]{David2004}%
  \BibitemOpen
  \bibfield  {author} {\bibinfo {author} {\bibfnamefont {D.~A.}\ \bibnamefont
  {Mazziotti}},\ }\bibfield  {title} {\enquote {\bibinfo {title} {Exactness of
  wave functions from two-body exponential transformations in many-body quantum
  theory},}\ }\href {\doibase 10.1103/PhysRevA.69.012507} {\bibfield  {journal}
  {\bibinfo  {journal} {Phys. Rev. A}\ }\textbf {\bibinfo {volume} {69}},\
  \bibinfo {pages} {012507} (\bibinfo {year} {2004})}\BibitemShut {NoStop}%
\bibitem [{\citenamefont {Garrod}\ and\ \citenamefont {Rosina}(1969)}]{GR1969}%
  \BibitemOpen
  \bibfield  {author} {\bibinfo {author} {\bibfnamefont {C.}~\bibnamefont
  {Garrod}}\ and\ \bibinfo {author} {\bibfnamefont {M.}~\bibnamefont
  {Rosina}},\ }\bibfield  {title} {\enquote {\bibinfo {title} {Particle‐hole
  matrix: its connection with the symmetries and collective features of the
  ground state},}\ }\href {\doibase 10.1063/1.1664770} {\bibfield  {journal}
  {\bibinfo  {journal} {J. Math. Phys.}\ }\textbf {\bibinfo {volume} {10}},\
  \bibinfo {pages} {1855--1861} (\bibinfo {year} {1969})}\BibitemShut {NoStop}%
\bibitem [{\citenamefont {Bose}\ and\ \citenamefont
  {Einstein}(1924)}]{bose_einstein_1924}%
  \BibitemOpen
  \bibfield  {author} {\bibinfo {author} {\bibfnamefont {S.~N.}\ \bibnamefont
  {Bose}}\ and\ \bibinfo {author} {\bibfnamefont {A.}~\bibnamefont
  {Einstein}},\ }\bibfield  {title} {\enquote {\bibinfo {title} {Planck's law
  and light quantum hypothesis},}\ }\href
  {http://web.ihep.su/dbserv/compas/src/bose24/eng.pdf} {\bibfield  {journal}
  {\bibinfo  {journal} {Zeitscrift für Physik}\ }\textbf {\bibinfo {volume}
  {26}},\ \bibinfo {pages} {178} (\bibinfo {year} {1924})}\BibitemShut
  {NoStop}%
\bibitem [{\citenamefont {Einstein}(1924)}]{einstein_1924}%
  \BibitemOpen
  \bibfield  {author} {\bibinfo {author} {\bibfnamefont {A.}~\bibnamefont
  {Einstein}},\ }\bibfield  {title} {\enquote {\bibinfo {title} {Quantentheorie
  des einatomigen idealen gases},}\ }\href
  {https://onlinelibrary.wiley.com/doi/abs/10.1002/3527608958.ch27} {\bibfield
  {journal} {\bibinfo  {journal} {K.P.A.W.}\ ,\ \bibinfo {pages} {261–267}}
  (\bibinfo {year} {1924})}\BibitemShut {NoStop}%
\bibitem [{\citenamefont {London}(1938)}]{london_1938}%
  \BibitemOpen
  \bibfield  {author} {\bibinfo {author} {\bibfnamefont {F.}~\bibnamefont
  {London}},\ }\bibfield  {title} {\enquote {\bibinfo {title} {On
  {B}ose-{E}instein condensation},}\ }\href {\doibase 10.1036/1097-8542.757474}
  {\bibfield  {journal} {\bibinfo  {journal} {Phys. Rev.}\ }\textbf {\bibinfo
  {volume} {54}},\ \bibinfo {pages} {947--–954} (\bibinfo {year}
  {1938})}\BibitemShut {NoStop}%
\bibitem [{\citenamefont {Tisza}(1947)}]{tisza_1947}%
  \BibitemOpen
  \bibfield  {author} {\bibinfo {author} {\bibfnamefont {L.}~\bibnamefont
  {Tisza}},\ }\bibfield  {title} {\enquote {\bibinfo {title} {The theory of
  liquid helium},}\ }\href@noop {} {\bibfield  {journal} {\bibinfo  {journal}
  {Phys. Rev.}\ }\textbf {\bibinfo {volume} {72}},\ \bibinfo {pages}
  {838--–854} (\bibinfo {year} {1947})}\BibitemShut {NoStop}%
\bibitem [{\citenamefont {Pauli}(1940)}]{pauli_1940}%
  \BibitemOpen
  \bibfield  {author} {\bibinfo {author} {\bibfnamefont {W.}~\bibnamefont
  {Pauli}},\ }\bibfield  {title} {\enquote {\bibinfo {title} {The connection
  between spin and statistics},}\ }\href {\doibase 10.1103/physrev.58.716}
  {\bibfield  {journal} {\bibinfo  {journal} {Physical Review}\ }\textbf
  {\bibinfo {volume} {58}},\ \bibinfo {pages} {716–722} (\bibinfo {year}
  {1940})}\BibitemShut {NoStop}%
\bibitem [{\citenamefont {Anderson}(2013)}]{Anderson_2013}%
  \BibitemOpen
  \bibfield  {author} {\bibinfo {author} {\bibfnamefont {P.~W.}\ \bibnamefont
  {Anderson}},\ }\bibfield  {title} {\enquote {\bibinfo {title} {Twenty-five
  years of high-temperature superconductivity {\textendash} a personal
  review},}\ }\href {\doibase 10.1088/1742-6596/449/1/012001} {\bibfield
  {journal} {\bibinfo  {journal} {J. Phys.: Conf. Ser.}\ }\textbf {\bibinfo
  {volume} {449}},\ \bibinfo {pages} {012001} (\bibinfo {year}
  {2013})}\BibitemShut {NoStop}%
\bibitem [{\citenamefont {Yang}(1962)}]{Y1962}%
  \BibitemOpen
  \bibfield  {author} {\bibinfo {author} {\bibfnamefont {C.~N.}\ \bibnamefont
  {Yang}},\ }\bibfield  {title} {\enquote {\bibinfo {title} {Concept of
  off-diagonal long-range order and the quantum phases of liquid {He} and of
  superconductors},}\ }\href {\doibase 10.1103/RevModPhys.34.694} {\bibfield
  {journal} {\bibinfo  {journal} {Rev. Mod. Phys.}\ }\textbf {\bibinfo {volume}
  {34}},\ \bibinfo {pages} {694--704} (\bibinfo {year} {1962})}\BibitemShut
  {NoStop}%
\bibitem [{\citenamefont {Sasaki}(1965)}]{S1965}%
  \BibitemOpen
  \bibfield  {author} {\bibinfo {author} {\bibfnamefont {F.}~\bibnamefont
  {Sasaki}},\ }\bibfield  {title} {\enquote {\bibinfo {title} {Eigenvalues of
  fermion density matrices},}\ }\href {\doibase 10.1103/PhysRev.138.B1338}
  {\bibfield  {journal} {\bibinfo  {journal} {Phys. Rev.}\ }\textbf {\bibinfo
  {volume} {138}},\ \bibinfo {pages} {B1338--B1342} (\bibinfo {year}
  {1965})}\BibitemShut {NoStop}%
\bibitem [{\citenamefont {Kohn}\ and\ \citenamefont
  {Sherrington}(1970)}]{Kohn1970}%
  \BibitemOpen
  \bibfield  {author} {\bibinfo {author} {\bibfnamefont {W.}~\bibnamefont
  {Kohn}}\ and\ \bibinfo {author} {\bibfnamefont {D.}~\bibnamefont
  {Sherrington}},\ }\bibfield  {title} {\enquote {\bibinfo {title} {Two kinds
  of bosons and bose condensates},}\ }\href {\doibase 10.1103/RevModPhys.42.1}
  {\bibfield  {journal} {\bibinfo  {journal} {Rev. Mod. Phys.}\ }\textbf
  {\bibinfo {volume} {42}},\ \bibinfo {pages} {1--11} (\bibinfo {year}
  {1970})}\BibitemShut {NoStop}%
\bibitem [{\citenamefont {Smart}\ \emph {et~al.}(2019)\citenamefont {Smart},
  \citenamefont {Schuster},\ and\ \citenamefont {Mazziotti}}]{smart_2019}%
  \BibitemOpen
  \bibfield  {author} {\bibinfo {author} {\bibfnamefont {S.~E.}\ \bibnamefont
  {Smart}}, \bibinfo {author} {\bibfnamefont {D.~I.}\ \bibnamefont {Schuster}},
  \ and\ \bibinfo {author} {\bibfnamefont {D.~A.}\ \bibnamefont {Mazziotti}},\
  }\bibfield  {title} {\enquote {\bibinfo {title} {Experimental data from a
  quantum computer verifies the generalized {P}auli exclusion principle},}\
  }\href {https://www.nature.com/articles/s42005-019-0110-3} {\bibfield
  {journal} {\bibinfo  {journal} {Communications Physics}\ }\textbf {\bibinfo
  {volume} {2}} (\bibinfo {year} {2019})}\BibitemShut {NoStop}%
\bibitem [{\citenamefont {Borland}\ and\ \citenamefont
  {Dennis}(1972)}]{borland_dennis_1972}%
  \BibitemOpen
  \bibfield  {author} {\bibinfo {author} {\bibfnamefont {R.~E.}\ \bibnamefont
  {Borland}}\ and\ \bibinfo {author} {\bibfnamefont {K.}~\bibnamefont
  {Dennis}},\ }\bibfield  {title} {\enquote {\bibinfo {title} {The conditions
  on the one-matrix for three-body fermion wavefunctions with one-rank equal to
  six},}\ }\href {\doibase 10.1088/0022-3700/5/1/009} {\bibfield  {journal}
  {\bibinfo  {journal} {J. Phys. B: At. Mol. Opt. Phys.}\ }\textbf {\bibinfo
  {volume} {5}},\ \bibinfo {pages} {7–15} (\bibinfo {year}
  {1972})}\BibitemShut {NoStop}%
\bibitem [{\citenamefont {Schilling}\ \emph {et~al.}(2013)\citenamefont
  {Schilling}, \citenamefont {Gross},\ and\ \citenamefont
  {Christandl}}]{schilling_2013}%
  \BibitemOpen
  \bibfield  {author} {\bibinfo {author} {\bibfnamefont {C.}~\bibnamefont
  {Schilling}}, \bibinfo {author} {\bibfnamefont {D.}~\bibnamefont {Gross}}, \
  and\ \bibinfo {author} {\bibfnamefont {M.}~\bibnamefont {Christandl}},\
  }\bibfield  {title} {\enquote {\bibinfo {title} {Pinning of fermionic
  occupation numbers},}\ }\href
  {https://pdfs.semanticscholar.org/3c9c/7391bb56738c63bc9cc7be21e5cdfeebc3c6.pdf?_ga=2.106787395.1978485703.1576770206-72123017.1576770206}
  {\bibfield  {journal} {\bibinfo  {journal} {Physical Review Letters}\
  }\textbf {\bibinfo {volume} {110}} (\bibinfo {year} {2013})}\BibitemShut
  {NoStop}%
\bibitem [{\citenamefont {Benavides-Riveros}\ \emph {et~al.}(2013)\citenamefont
  {Benavides-Riveros}, \citenamefont {Gracia-Bondía},\ and\ \citenamefont
  {Springborg}}]{benavides_2013}%
  \BibitemOpen
  \bibfield  {author} {\bibinfo {author} {\bibfnamefont {C.~L.}\ \bibnamefont
  {Benavides-Riveros}}, \bibinfo {author} {\bibfnamefont {J.~M.}\ \bibnamefont
  {Gracia-Bondía}}, \ and\ \bibinfo {author} {\bibfnamefont {M.}~\bibnamefont
  {Springborg}},\ }\bibfield  {title} {\enquote {\bibinfo {title} {Quasipinning
  and entanglement in the lithium isoelectronic series},}\ }\href
  {https://journals.aps.org/pra/abstract/10.1103/PhysRevA.88.022508} {\bibfield
   {journal} {\bibinfo  {journal} {Physical Review A}\ }\textbf {\bibinfo
  {volume} {88}} (\bibinfo {year} {2013})}\BibitemShut {NoStop}%
\bibitem [{\citenamefont {Chakraborty}\ and\ \citenamefont
  {Mazziotti}(2014)}]{chakraborty_2014}%
  \BibitemOpen
  \bibfield  {author} {\bibinfo {author} {\bibfnamefont {R.}~\bibnamefont
  {Chakraborty}}\ and\ \bibinfo {author} {\bibfnamefont {D.~A.}\ \bibnamefont
  {Mazziotti}},\ }\bibfield  {title} {\enquote {\bibinfo {title} {Generalized
  pauli conditions on the spectra of one-electron reduced density matrices of
  atoms and molecules},}\ }\href
  {https://journals.aps.org/pra/abstract/10.1103/PhysRevA.89.042505} {\bibfield
   {journal} {\bibinfo  {journal} {Physical Review A}\ }\textbf {\bibinfo
  {volume} {89}} (\bibinfo {year} {2014})}\BibitemShut {NoStop}%
\bibitem [{\citenamefont {Mazziotti}(2016)}]{mazziotti_2016}%
  \BibitemOpen
  \bibfield  {author} {\bibinfo {author} {\bibfnamefont {D.~A.}\ \bibnamefont
  {Mazziotti}},\ }\bibfield  {title} {\enquote {\bibinfo {title}
  {Pure-{N}-representability conditions of two-fermion reduced density
  matrices},}\ }\href
  {https://journals.aps.org/pra/abstract/10.1103/PhysRevA.94.032516} {\bibfield
   {journal} {\bibinfo  {journal} {Physical Review A}\ }\textbf {\bibinfo
  {volume} {94}} (\bibinfo {year} {2016})}\BibitemShut {NoStop}%
\bibitem [{\citenamefont {Walter}\ \emph {et~al.}(2016)\citenamefont {Walter},
  \citenamefont {Gross},\ and\ \citenamefont {Eisert}}]{wge_2016}%
  \BibitemOpen
  \bibfield  {author} {\bibinfo {author} {\bibfnamefont {M.}~\bibnamefont
  {Walter}}, \bibinfo {author} {\bibfnamefont {D.}~\bibnamefont {Gross}}, \
  and\ \bibinfo {author} {\bibfnamefont {J.}~\bibnamefont {Eisert}},\
  }\bibfield  {title} {\enquote {\bibinfo {title} {Multipartite
  entanglement},}\ }\href {\doibase 10.1002/9783527805785.ch14} {\bibfield
  {journal} {\bibinfo  {journal} {Quantum Information}\ ,\ \bibinfo {pages}
  {293–330}} (\bibinfo {year} {2016})}\BibitemShut {NoStop}%
\bibitem [{\citenamefont {Cruz}\ \emph {et~al.}(2019)\citenamefont {Cruz},
  \citenamefont {Fournier}, \citenamefont {Gremion}, \citenamefont {Jeannerot},
  \citenamefont {Komagata}, \citenamefont {Tosic}, \citenamefont
  {Thiesbrummel}, \citenamefont {Chan}, \citenamefont {Macris}, \citenamefont
  {Dupertuis},\ and\ \citenamefont {Javerzac-Galy}}]{cruz_2019}%
  \BibitemOpen
  \bibfield  {author} {\bibinfo {author} {\bibfnamefont {D.}~\bibnamefont
  {Cruz}}, \bibinfo {author} {\bibfnamefont {R.}~\bibnamefont {Fournier}},
  \bibinfo {author} {\bibfnamefont {F.}~\bibnamefont {Gremion}}, \bibinfo
  {author} {\bibfnamefont {A.}~\bibnamefont {Jeannerot}}, \bibinfo {author}
  {\bibfnamefont {K.}~\bibnamefont {Komagata}}, \bibinfo {author}
  {\bibfnamefont {T.}~\bibnamefont {Tosic}}, \bibinfo {author} {\bibfnamefont
  {J.}~\bibnamefont {Thiesbrummel}}, \bibinfo {author} {\bibfnamefont {C.~L.}\
  \bibnamefont {Chan}}, \bibinfo {author} {\bibfnamefont {N.}~\bibnamefont
  {Macris}}, \bibinfo {author} {\bibfnamefont {M.-A.}\ \bibnamefont
  {Dupertuis}}, \ and\ \bibinfo {author} {\bibfnamefont {C.}~\bibnamefont
  {Javerzac-Galy}},\ }\bibfield  {title} {\enquote {\bibinfo {title} {Efficient
  quantum algorithms for ghz and w states, and implementation on the ibm
  quantum computer},}\ }\href
  {https://onlinelibrary.wiley.com/doi/abs/10.1002/qute.201900015} {\bibfield
  {journal} {\bibinfo  {journal} {Advanced Quantum Technologies}\ }\textbf
  {\bibinfo {volume} {2}},\ \bibinfo {pages} {1900015} (\bibinfo {year}
  {2019})}\BibitemShut {NoStop}%
\bibitem [{\citenamefont {Treinish}\ and\ \citenamefont
  {Rodríguez}()}]{TR_GHZ}%
  \BibitemOpen
  \bibfield  {author} {\bibinfo {author} {\bibfnamefont {M.}~\bibnamefont
  {Treinish}}\ and\ \bibinfo {author} {\bibfnamefont {D.~M.}\ \bibnamefont
  {Rodríguez}},\ }\bibfield  {title} {\enquote {\bibinfo {title} {{GHZ} state
  example},}\ }\href
  {https://github.com/Qiskit/qiskit-terra/blob/master/examples/python/ibmq/ghz.py}
  {\bibinfo  {journal} {GitHub}\ }\BibitemShut {NoStop}%
\bibitem [{\citenamefont {IBM-Q-Team}(2019{\natexlab{a}})}]{qx2_2019}%
  \BibitemOpen
\bibfield  {journal} {  }\bibfield  {author} {\bibinfo {author} {\bibnamefont
  {IBM-Q-Team}},\ }\href@noop {} {\enquote {\bibinfo {title} {{IBM-Q-5}
  {Y}orktown backend specification v2.0.0},}\ } (\bibinfo {year}
  {2019}{\natexlab{a}})\BibitemShut {NoStop}%
\bibitem [{\citenamefont {IBM-Q-Team}(2019{\natexlab{b}})}]{melbourne}%
  \BibitemOpen
  \bibfield  {author} {\bibinfo {author} {\bibnamefont {IBM-Q-Team}},\
  }\href@noop {} {\enquote {\bibinfo {title} {{IBM-Q-15} {M}elbourne backend
  specification v2.0.0},}\ } (\bibinfo {year} {2019}{\natexlab{b}})\BibitemShut
  {NoStop}%
\bibitem [{\citenamefont {IBM-Q-Team}(2020)}]{rochester}%
  \BibitemOpen
  \bibfield  {author} {\bibinfo {author} {\bibnamefont {IBM-Q-Team}},\
  }\href@noop {} {\enquote {\bibinfo {title} {{IBM-Q-53} {R}ochester backend
  specification v1.2.0},}\ } (\bibinfo {year} {2020})\BibitemShut {NoStop}%
\bibitem [{\citenamefont {Koch}\ \emph {et~al.}(2007)\citenamefont {Koch},
  \citenamefont {Yu}, \citenamefont {Gambetta}, \citenamefont {Houck},
  \citenamefont {Schuster}, \citenamefont {Majer}, \citenamefont {Blais},
  \citenamefont {Devoret}, \citenamefont {Girvin}, \citenamefont {Schoelkopf},\
  and\ \citenamefont {et. al.}}]{koch_2007}%
  \BibitemOpen
  \bibfield  {author} {\bibinfo {author} {\bibfnamefont {J.}~\bibnamefont
  {Koch}}, \bibinfo {author} {\bibfnamefont {T.~M.}\ \bibnamefont {Yu}},
  \bibinfo {author} {\bibfnamefont {J.}~\bibnamefont {Gambetta}}, \bibinfo
  {author} {\bibfnamefont {A.~A.}\ \bibnamefont {Houck}}, \bibinfo {author}
  {\bibfnamefont {D.~I.}\ \bibnamefont {Schuster}}, \bibinfo {author}
  {\bibfnamefont {J.}~\bibnamefont {Majer}}, \bibinfo {author} {\bibfnamefont
  {A.}~\bibnamefont {Blais}}, \bibinfo {author} {\bibfnamefont {M.~H.}\
  \bibnamefont {Devoret}}, \bibinfo {author} {\bibfnamefont {S.~M.}\
  \bibnamefont {Girvin}}, \bibinfo {author} {\bibfnamefont {R.~J.}\
  \bibnamefont {Schoelkopf}}, \ and\ \bibinfo {author} {\bibnamefont {et.
  al.}},\ }\bibfield  {title} {\enquote {\bibinfo {title} {Charge-insensitive
  qubit design derived from the cooper pair box},}\ }\href
  {https://journals.aps.org/pra/abstract/10.1103/PhysRevA.76.042319} {\bibfield
   {journal} {\bibinfo  {journal} {Physical Review A}\ }\textbf {\bibinfo
  {volume} {76}} (\bibinfo {year} {2007})}\BibitemShut {NoStop}%
\bibitem [{\citenamefont {Chow}\ \emph {et~al.}(2011)\citenamefont {Chow},
  \citenamefont {Córcoles}, \citenamefont {Gambetta}, \citenamefont {Rigetti},
  \citenamefont {Johnson}, \citenamefont {Smolin}, \citenamefont {Rozen},
  \citenamefont {Keefe}, \citenamefont {Rothwell}, \citenamefont {Ketchen},\
  and\ \citenamefont {et~al.}}]{chow_2011}%
  \BibitemOpen
  \bibfield  {author} {\bibinfo {author} {\bibfnamefont {J.~M.}\ \bibnamefont
  {Chow}}, \bibinfo {author} {\bibfnamefont {A.~D.}\ \bibnamefont {Córcoles}},
  \bibinfo {author} {\bibfnamefont {J.~M.}\ \bibnamefont {Gambetta}}, \bibinfo
  {author} {\bibfnamefont {C.}~\bibnamefont {Rigetti}}, \bibinfo {author}
  {\bibfnamefont {B.~R.}\ \bibnamefont {Johnson}}, \bibinfo {author}
  {\bibfnamefont {J.~A.}\ \bibnamefont {Smolin}}, \bibinfo {author}
  {\bibfnamefont {J.~R.}\ \bibnamefont {Rozen}}, \bibinfo {author}
  {\bibfnamefont {G.~A.}\ \bibnamefont {Keefe}}, \bibinfo {author}
  {\bibfnamefont {M.~B.}\ \bibnamefont {Rothwell}}, \bibinfo {author}
  {\bibfnamefont {M.~B.}\ \bibnamefont {Ketchen}}, \ and\ \bibinfo {author}
  {\bibnamefont {et~al.}},\ }\bibfield  {title} {\enquote {\bibinfo {title}
  {Simple all-microwave entangling gate for fixed-frequency superconducting
  qubits},}\ }\href
  {https://journals.aps.org/prl/abstract/10.1103/PhysRevLett.107.080502}
  {\bibfield  {journal} {\bibinfo  {journal} {Physical Review Letters}\
  }\textbf {\bibinfo {volume} {107}} (\bibinfo {year} {2011})}\BibitemShut
  {NoStop}%
\end{thebibliography}%

\end{document}